\documentclass[journal]{IEEEtran}
\usepackage{amsmath,amsfonts}
\usepackage{algorithmic}
\usepackage{algorithm}
\usepackage{array}
\usepackage{textcomp}
\usepackage{stfloats}
\usepackage{url}
\usepackage{verbatim}
\usepackage{graphicx}
\usepackage{cite}
\usepackage{calligra}
\usepackage{makecell}
\usepackage{xcolor}
\usepackage{bm} 
\usepackage{subfigure}
\usepackage{amsmath,amsfonts}
\usepackage{multirow}

\begin{document}

\title{Message Passing-Based Joint Channel Estimation and Signal Detection for OTFS with Superimposed Pilots}

\author{Fupeng Huang, Qinghua Guo, {\em Senior Member, IEEE}, Youwen Zhang and Yuriy V. Zakharov, {\em Senior Member, IEEE}
\thanks{Fupeng Huang is with National Key Laboratory of underwater Acoustic Technology, Harbin Engineering University, Harbin 150001, China, and Key Laboratory of Marine Information Acquisition and Security (Harbin Engineering University), Ministry of Industry and Information Technology, Harbin 150001, China, and College of Underwater Acoustic Engineering, Harbin Engineering University, Harbin 150001, China. (corresponding author, e-mail: huangfupeng@126.com).}
\thanks{Qinghua Guo is with the School of Electrical, Computer and Telecommunications Engineering, University of Wollongong, Wollongong, NSW 2522, Australia. (e-mail: qguo@uow.edu.au)}
\thanks{Youwen Zhang is with National Key Laboratory of underwater Acoustic Technology, Harbin Engineering University, Harbin 150001, China, and Key Laboratory of Marine Information Acquisition and Security (Harbin Engineering University), Ministry of Industry and Information Technology, Harbin 150001, China, and College of Underwater Acoustic Engineering, Harbin Engineering University, Harbin 150001, China. He is also with Ocean College, Jiangsu University of Science and Technology, Zhenjiang, 212100, China. (e-mail: zhangyuwnjust@163.com)}
\thanks{Yuriy V. Zakharov is with the School of Physics, Engineering and Technology, University of York, York, YO10 5DD, U.K. The work of Y. Zakharov was supported in part by the U.K. Engineering and Physical Sciences Research Council (EPSRC) through Grants EP/V009591/1 and EP/R003297/1. (e-mail: yury.zakharov@york.ac.uk) }}


\IEEEpubid{\begin{minipage}{\textwidth}\ \centering
		Copyright \copyright 2015 IEEE. Personal use of this material is permitted. \\
		However, permission to use this material for any other purposes must be obtained from the IEEE by sending a request to pubs-permissions@ieee.org.
\end{minipage}}

\maketitle
\begin{abstract}
Receivers with joint channel estimation (CE) and signal detection using superimposed pilots (SP) can achieve high transmission efficiency in orthogonal time frequency space (OTFS) systems. However, existing receivers have high computational complexity, hindering their practical applications. In this work, with SP in the delay-Doppler (DD) domain and the generalized complex exponential (GCE) basis expansion modeling (BEM) for channels, a message passing-based SP-DD iterative receiver is proposed, which drastically reduces the computational complexity while with marginal performance loss, compared to existing ones. To facilitate CE in the proposed receiver, we design pilot signal to achieve pilot power concentration in the frequency domain, thereby developing an SP-DD-D receiver that can reduce the power of the pilot signal with almost no loss of CE and bit error rate (BER) performance. Extensive simulation results are provided to demonstrate the superiority of the proposed SP-DD-D receiver.

\end{abstract}


\section{Introduction}
Orthogonal time frequency space (OTFS) modulation in the delay-Doppler (DD) domain has attracted recently much attention, due to its capability of achieving reliable communications in high mobility applications (thanks to its full time and frequency diversity) \cite{Hadani2017,  Raviteja2018, Raviteja2019, Surabhi2019, Farhang2018, Hadani2018}. To unleash the full potentials of OTFS, a practical and powerful receiver with joint channel estimation (CE) and signal detection is essential, which is the focus of this paper.

With the assumption that the channel state information (CSI) is known, various OTFS signal detectors have been designed. A maximum-likelihood (ML) detector in multiple input multiple output (MIMO) OTFS system was designed in \cite{Surabhi2019}, which can achieve full diversity. A low complexity two-stage detector was proposed in \cite{Li2017online}, where equalization is performed in the Doppler domain through pre-processing in the frequency domain. In \cite{Thaj2020}, a low complexity iterative Rake decision feedback detector was proposed, where the maximum ratio combining (MRC) is used to improve the SNR of the combined signal. Linear minimum mean squared error (LMMSE) and zero-forcing (ZF) detectors with low-complexity were proposed in \cite{Surabhi2020}, where a local search technique is used to achieve low-complexity initial solutions. In \cite{Raviteja2018}, \cite{Mishra2017}, \cite{Liu2022} and \cite{Liu2022a}, message passing (MP) based OTFS receivers were developed, which implement a lower complexity OTFS receiver with the message passing techniques. A variational Bayes (VB) detector was proposed in \cite{Yuan2020TVT} to achieve better convergence compared with the existing message passing based detectors. In \cite{Guo2022TWC} and \cite{Guo2022LWC}, (unitary) approximate message passing (AMP) based detectors were designed, which show outstanding performance while with low complexity. A linear MMSE based parallel interference cancellation (LMMSE-PIC) equalization was proposed in \cite{Long2019}, which uses the first order Neumann series to reduce the complexity. In \cite{Long2021} and \cite{Long2022}, the expectation propagation (EP) was used to implement an OTFS detector, which shows superior performance compared to the LMMSE receiver. In \cite{Li2021TVT}, a hybrid detection algorithm was proposed. It uses a partitioning rule to divide the relevant received symbols into two subsets to detect each transmitted symbol, where the maximum $a$ $posteriori$ (MAP) detection is applied to the subset with larger channel gains, and the parallel interference cancellation (PIC) detection is applied to the subset with smaller channel gains. These signal detectors assume the exact knowledge of CSI, which has to acquired through CE. In addition, CE errors are ignored in the design of the detectors.

\IEEEpubidadjcol

CE schemes in OTFS systems can be categorized into three types, such as the full pilot scheme, embedded pilot scheme, and superimposed pilot (SP) scheme. In the full pilot scheme, a frame comprising only one non-zero pilot symbol \cite{Ramachandran2013} is dedicated to CE, and the estimated channel is used for detection in subsequent frames. The full pilot scheme leads to low spectrum efficiency, but also to difficulties in dealing with fast time-varying channels. To overcome these problems, the embedded pilot scheme is used to estimate the DD domain channel, e.g., in \cite{Raviteja2019}, \cite{Liu2022}, \cite{Long2021} and \cite{Guo2021TWC}, where CE and signal detection are performed at the same frame. Although the embedded pilot scheme can deal with fast time-varying channels, the problem of low spectral efficiency is still a concern due to the use of guard interval between pilots and data, especially for long delay and/or high Doppler shift channels. In order to improve the spectral efficiency, the SP scheme is promising, where the pilots are superimposed with data and guard intervals are no longer needed \cite{Mishra2017}, \cite{Liu2022a}, \cite{Guo2009TWC, Budianu2002, Tugnait2003,GuoTSP2011, Carrasco-Alvarez2012, Yuan2021}. Moreover, to reduce the complexity of the SP scheme, the basis expansion modeling (BEM) has been applied in \cite{Liu2022}, \cite{Liu2022a} and \cite{Long2021}, which can significantly reduce the number of unknown channel parameters. There are many types of BEMs, such as complex exponential BEM (CE-BEM), generalized CE-BEM (GCE-BEM), discrete prolate spheroidal BEM (DPS-BEM), Karhunen-Loeve BEM (KL-BEM), etc  \cite{Cheng2013, Tang2017, He2008, Long2021, Liu2022, Liu2022a}. The SP scheme is attractive in terms of spectral efficiency and dealing with fast time-varying channels, but joint CE and signal detection need to be performed to deal with the interference between pilots and data. The existing OTFS receivers with the SP scheme require high-dimensional matrix operations and inverse (or pseudo inverse) operations \cite{Mishra2017}, \cite{Liu2022a}, which is a serious concern due to the high computational complexity involved.

In this paper, we aim to develop a low complexity OTFS receiver with the SP scheme. To reduce the number of unknown channel parameters and improve the accuracy of CE, we adopt the GCE-BEM CE. In this work, leveraging the message passing techniques, with SP in the DD domain and the GCE BEM for channels, a message passing based iterative receiver called the SP-DD receiver is proposed. Compared to existing receivers, the SP-DD receiver drastically reduces the computational complexity while with marginal performance loss. To make the SP scheme more efficient, we design the pilot signals carefully to achieve pilot power concentration, facilitating the reduction of the pilot power without decreasing the performance of the receiver, leading to a receiver named SP-DD-D, which can also reduce the peak-to-average power ratio (PAPR) of OTFS signals. Extensive simulation results are provided to demonstrate the superiority of the proposed receivers against existing receivers.

This paper is organized as follows. In Section \ref{Sys-M}, the OTFS system model is presented. In Section \ref{MPFG}, leveraging the message passing techniques, we first propose the SP-DD receiver, and then, with carefully designed pilot signals, the SP-DD-D receiver is proposed. Pilot power ratio determination and complexity analysis are given in Section \ref{PA}. Simulation results are provided in Section \ref{SR} to demonstrate the performance of the proposed receivers, and the paper is concluded in Section \ref{sec7}.

\emph{Notation:}
Matrices and vectors are denoted by uppercase and lowercase bold letters, respectively. The identity matrix is represented as ${\bf{I}}_M$ (${\bf{I}}_M \in {\bf{C}}^{M \times M}$). The notation $\otimes$ denotes the Kronecker product, and $\odot$ and $./$ represent the element-wise product and division operations. We use $diag\left( {{a_1},{a_2}, \cdots ,{a_n}} \right)$ to represent a diagonal matrix with the diagonal elements ${a_1},{a_2}, \cdots ,{a_n}$. The operator ${\bf{vec}}(\bf{A})$ reshapes the matrix $\bf{A}$ to produce a column vector, and ${\bf{vec}^{-1}}(\bf{a})$ is the inverse operation of ${\bf{vec}}(\bf{A})$, which returns the matrix $\bf{A}$. The notation $\lceil\cdot\rceil$ represents the rounding up operation, and $<\cdot>$ represents the averaging operation.

\section{OTFS Transmitter and Channel Model}\label{Sys-M}
\subsection{Transmitter Structure}
\begin{figure}[ht]
	\centering
	\includegraphics[scale=.45]{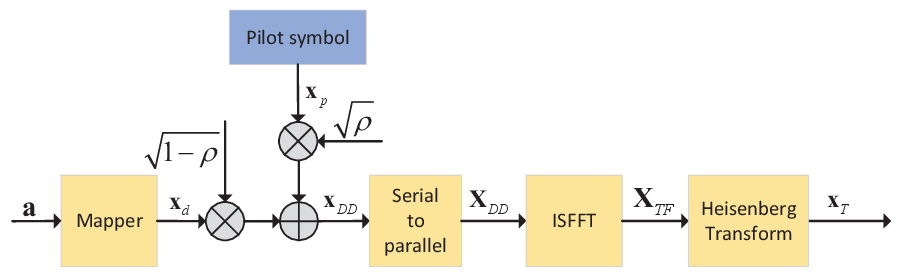}
	\caption{Structure of the OTFS transmitter with the SP scheme.}
	\label{FIG:TXstructure}
\end{figure}
The structure of the OTFS transmitter with the SP scheme is shown in Fig. \ref{FIG:TXstructure}, where ${\bf{a}}$ denotes an information bit sequence, and every $K$  bits of ${\bf{a}}$ are mapped into a symbol in a ${2^K}$-ary constellation set $A = \left\{ {{\alpha _1},  \cdots,  {\alpha _{{2^K}}}} \right\}$, resulting in a symbol ${\bf{x}}_d$. The data symbols and the pilot symbols are given by
\begin{equation}
{\bf{x}}_d = \left\{x_d[0], x_d[1], \cdots , x_d[MN-1]\right\}^T,
\label{eq:x_d}
\end{equation}
\begin{equation}
{\bf{x}}_p = \left\{x_p[0], x_p[1], \cdots , x_p[MN-1]\right\}^T,
\label{eq:x_p}
\end{equation}
where $M$ and $N$ denote the number of delay bins and Doppler bins, respectively. Thus the symbol sequence in the DD domain ${\bf{x}}_{DD}$ can be expressed as \cite{Liu2022a}

\begin{equation}
{\bf{x}}_{DD} = \sqrt{\rho}{\bf{x}}_p + \sqrt{1 - \rho}{\bf{x}}_d,
\label{eq:xDD}
\end{equation}
where $0<\rho<1$ is a power allocation factor. We define a DD domain symbol matrix ${\bf{X}}_{DD} = {\bf{vec}^{-1}}({\bf{x}}_{DD})$ and ${\bf{X}}_{DD} \in {\bf{C}}^{M \times N}$. After the inverse symplectic finite Fourier transform (ISFFT), the signals in the time-frequency (TF) domain are given as \cite{Mishra2017}
\begin{equation}
{\bf{X}}_{TF} = {\bf{F}}_{M}{\bf{X}}_{DD}{{\bf{F}}^H_N},
\label{eq:XTF}
\end{equation}
where ${\bf{F}}_{M} \in {\bf{C}}^{M \times M}$ and ${\bf{F}}_{N} \in {\bf{C}}^{N \times N}$ are normalized discrete Fourier transform (DFT) matrices with ${\bf{F}}_{M}(p, q) = \sqrt{1/M}\exp(-j2\pi pq/M)$ and
${\bf{F}}_{N}(p, q) = \sqrt{1/N}\exp(-j2\pi pq/N)$.
After the Heisenberg transform, the time-domain signal can be represented as
\begin{equation}
{\bf{x}}_{T} = {\bf{vec}}\left({{\bf{F}}^H_M}{\bf{X}}_{TF}\right) = {\bf{vec}}\left({\bf{X}}_{DD}{{\bf{F}}^H_N}\right).
\label{eq:x_T}
\end{equation}
As $\bf{vec}({\bf ABC}) = \left({\bf C^T} \otimes {\bf A}\right){\bf vec}\left({\bf B}\right)$ \cite{Mishra2017}, the transmitted signal (\ref{eq:x_T}) can be rewritten as
\begin{equation}
{\bf{x}}_{T} = \left({\bf{F}}^{H}_N \otimes {\bf{I}}_M\right)\left(\sqrt{\rho}{\bf{x}}_p + \sqrt{1 - \rho}{\bf{x}}_d\right).
\label{eq:x_T1}
\end{equation}
\subsection{Channel Model}\label{C-M}
We assume that the channel memory length is $L$. After removing the cyclic prefix (CP), the received signal can be expressed as
\begin{equation}
{\bf{y}}_{T} = {\bf{H}}_T{\bf{x}}_{T} + {\bf{w}},
\label{eq:t2r}
\end{equation}
where $\bf{w}$ is the additive white Gaussian noise (AWGN), and ${\bf{H}}_T$ is the time domain channel matrix defined as \cite{Liu2022a} (at the
top of next page).
\begin{figure*}[htbp]
\begin{equation}
{{\bf{H}}_T} = \left[ {\begin{array}{*{10}{c}}
h(0, 0)&  \cdots&     0& h(0, L-1)&  h(0, L-2)& \cdots& h(0, 1)\\
h(1, 1)& h(1, 0)& \cdots&         0&  h(1, L-1)& \cdots& h(1, 2)\\
\vdots & \vdots&\ddots&    \ddots&     \vdots& \vdots& \vdots \\
      0&  \cdots&  0& h(MN, L-1)&  h(MN, L-2)& \cdots& h(MN, 0)\\
\end{array}} \right].
\label{eq:HT}
\end{equation}
\end{figure*}
Substituting (\ref{eq:x_T1}) into (\ref{eq:t2r}), we obtain
\begin{equation}
{\bf{y}}_{T} = {\bf{H}}_T\left({\bf{F}}^{H}_N \otimes {\bf{I}}_M\right) \left(\sqrt{\rho}{\bf{x}}_p + \sqrt{1 - \rho}{\bf{x}}_d\right)+ {\bf{w}}.
\label{eq:t2r_TDD}
\end{equation}
With the BEM, ${\bf{H}}_T$ in (\ref{eq:HT}) can be modeled as \cite{Liu2022, Liu2022a}
\begin{equation}
{\bf{H}}_T = \sum_{q = 0}^{Q-1}diag\{{{\bf{b}}_q}\}{\bf{C}}_q + {\bf{E}}_m,
\label{eq:BEM_Ht}
\end{equation}
where ${{\bf{b}}_q}$ is an GCE BEM basis function, $Q$ denotes the order of the BEM basis, ${\bf{E}}_m$ is the channel modeling error matrix, and ${\bf{C}}_q$ is a circulant matrix, which can be expressed as \cite{Liu2022, Liu2022a}
\begin{equation}
\begin{split}
{\bf{C}}_q = \sqrt{MN}{\bf{F}}^{H}_{MN}diag\{ {\bf{F}}_{MN \times L}{\bf{c}}^{'}_q\}{\bf{F}}_{MN} \\
= {\bf{F}}^{H}_{MN}diag\{ {\bf{F}}_{MN \times L}{\bf{c}}_q\}{\bf{F}}_{MN},
\label{eq:BEM_Cq}
\end{split}
\end{equation}
where ${\bf{c}}_q = \sqrt{MN}{\bf{c}}^{'}_q$ denotes the $qth$ BEM cofficient, and ${\bf{F}}_{MN \times L}$corresponds to the first $L$ columns of ${\bf{F}}_{MN}$. Substituting (\ref{eq:BEM_Cq}),  (\ref{eq:BEM_Ht}) into (\ref{eq:t2r_TDD}), we can obtain the received signal
\begin{flalign}
\begin{split}
{\bf{y}}_{T} \approx \sum_{q = 0}^{Q-1}diag\{{{\bf{b}}_q}\}{\bf{F}}^{H}_{MN}diag\{ {\bf{F}}_{MN \times L}{\bf{c}}_q\}{\bf{F}}_{MN}\\
({\bf{F}}^{H}_N \otimes {\bf{I}}_M) \left(\sqrt{\rho}{\bf{x}}_p + \sqrt{1 - \rho}{\bf{x}}_d \right)+ {\bf{w}}.
\label{eq:t2rBEMTDD}
\end{split}
\end{flalign}

In this work, with the superimposed pilots, we aim to jointly estimate the channel and perform signal detection. In particular, we will use the Bayesian approach and implement the receiver using the factor graph and message passing techniques, so that the joint CE and signal detection can be achieved in an iterative manner. In addition, as superimposed pilot scheme is used, the pilot overhead for channel estimation is due to the power occupied by the pilots, i.e., the power overhead is give as $10log(1 - \rho)$ dB where $\rho$ is the power ration of the pilots.

\section{Proposed Message Passing Based Receivers}\label{MPFG}

In this section, we will design a message passing based Bayesian receiver. To achieve this, we need to find the joint distribution of the relevant variables, and construct a factor graph representation. Then message passing algorithms can be developed based on the factor graph.

\subsection{Factor Graph Representation}
\begin{figure*}[hbpt]
	\centering
	\includegraphics[scale=.45]{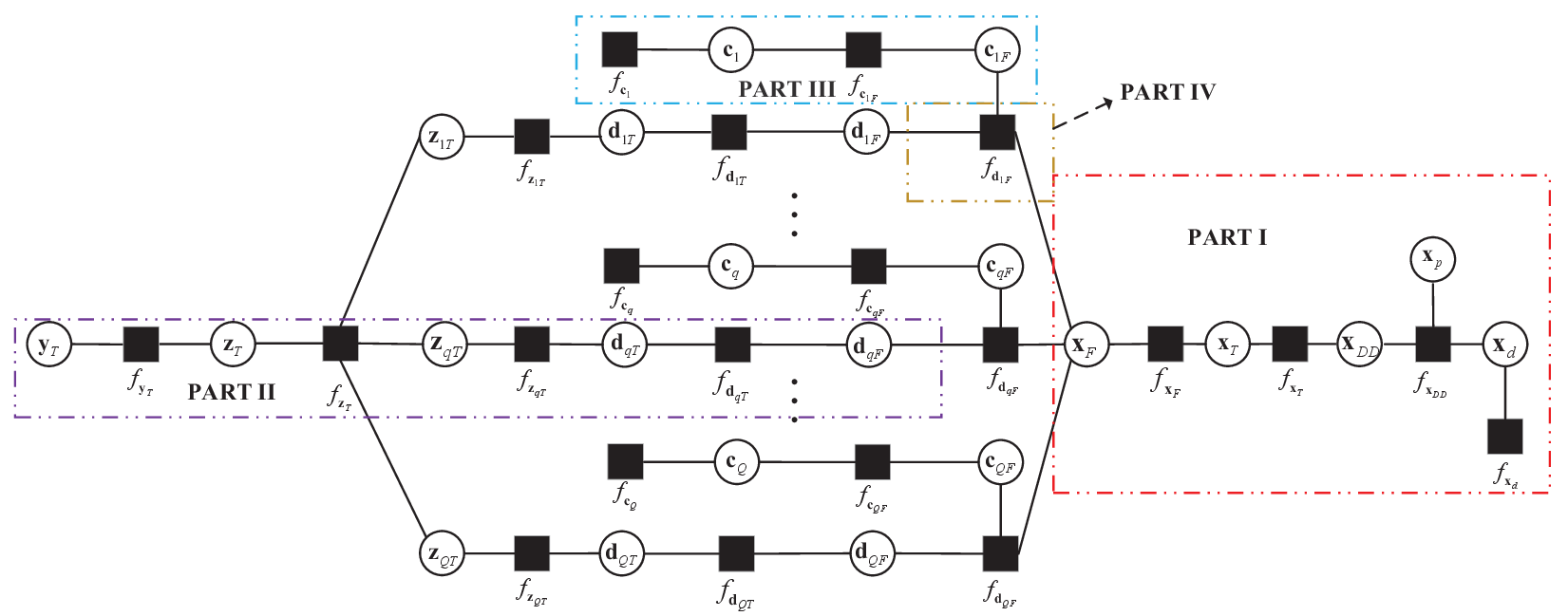}
	\caption{Factor graph representation for developing the OTFS receivers.}
	\label{FIG:RXstructure}
\end{figure*}

For the convenience of the algorithm design, we define the following auxiliary variables,
\begin{equation}
{\bf{x}}_{F} = {\bf{F}}_{MN} {\bf{x}}_{T},
\label{eq:x_F}
\end{equation}
\begin{equation}
{\bf{c}}_{qF} = {\bf{F}}_{MN \times L} {\bf{c}}_{q},
\label{eq:c_qF}
\end{equation}
\begin{equation}
{\bf{d}}_{qF} = {\bf{x}}_{F} \odot {\bf{c}}_{qF},
\label{eq:d_qF}
\end{equation}
\begin{equation}
{\bf{d}}_{qT} = {\bf{F}}^{H}_{MN} {\bf{d}}_{qF},
\label{eq:d_T}
\end{equation}
\begin{equation}
{\bf{z}}_{qT} = {\bf{b}}_{q} \odot {\bf{d}}_{qT},
\label{eq:z_qT}
\end{equation}
\begin{equation}
{\bf{z}}_{T} = \sum_{q=0}^{Q-1} {\bf{z}}_{qT},
\label{eq:z_T}
\end{equation}
\begin{equation}
{\bf{y}}_{T} = {\bf{z}}_{T} + {\bf{w}}.
\label{eq:y_T}
\end{equation}

With these variables, it is not hard to get the following joint conditional distribution and its factorization
\begin{flalign}
\begin{split}
&p\left({\bf{x}}_d, {\bf{x}}_{DD}, {\bf{x}}_{T}, {\bf{x}}_{F}, {\bf{c}}_q, {\bf{c}}_{qF}, {\bf{d}}_{qF}, {\bf{d}}_{qT}, {\bf{z}}_{qT}, {\bf{z}}_T|{\bf{y}}_T\right)     \\
&\propto p\left({\bf{y}}_{T}|{\bf{z}}_{T}\right)p\left({\bf{z}}_{T}|{\bf{z}}_{qT}\right)p\left({\bf{z}}_{qT}|{\bf{d}}_{qT}\right)p\left({\bf{d}}_{qT}|{\bf{d}}_{qF}\right)p\left({\bf{d}}_{qF}|{\bf{c}}_{qF}, {\bf{x}}_{F}\right)\\
&p\left({\bf{c}}_{qF}|{\bf{c}}_{q}\right)p\left({\bf{x}}_{F}|{\bf{x}}_{T}\right)p\left({\bf{x}}_{T}|{\bf{x}}_{DD}\right)p\left({\bf{x}}_{DD}|{\bf{x}}_{d}\right)p\left({\bf{x}}_{d}\right) \\
& = f_{{\bf{y}}_{T}}\left({\bf{y}}_{T}, {\bf{z}}_{T}\right)f_{{\bf{z}}_{T}}\left({\bf{z}}_{T},{\bf{z}}_{qT}\right)f_{{\bf{z}}_{qT}}\left({\bf{z}}_{qT},{\bf{d}}_{qT}\right)f_{{\bf{d}}_{qT}}\left({\bf{d}}_{qT},{\bf{d}}_{qF}\right)\\
&f_{{\bf{d}}_{qF}}\left({\bf{d}}_{qF}, {\bf{c}}_{qF},{\bf{x}}_{F}\right)f_{{\bf{c}}_{qF}}\left({\bf{c}}_{qF},{\bf{c}}_{q}\right)f_{{\bf{c}}_{q}}\left({\bf{c}}_{q}\right)f_{{\bf{x}}_{F}}\left({\bf{x}}_{F}, {\bf{x}}_{T}\right)\\
&f_{{\bf{x}}_{T}}\left({\bf{x}}_{T}, {\bf{x}}_{DD}\right)f_{{\bf{x}}_{DD}}\left({\bf{x}}_{DD}, {\bf{x}}_{d}\right)f_{{\bf{x}}_{d}}\left({\bf{x}}_{d}\right).
\label{eq:joint_cd}
\end{split}
\end{flalign}

\begin{table}[htbp]
 \caption{\label{tbl0}Factors and distributions}
 \renewcommand\arraystretch{1.5}
 \begin{center}
 \begin{tabular}{l|l|l}
  \hline {\bf{Factor}} & {\bf{Distribution}}  & {\bf{Function}} \\
  \hline   $f_{{\bf{y}}_{T}}$ & $p\left({\bf{y}}_{T}|{\bf{z}}_{T}\right)$ & $N\left({\bf{z}}_{T};{\bf{y}}_{T}, w^{-1}{\bf{I}}\right)$\\
  \hline   $f_{{\bf{z}}_{T}}$ & $p\left({\bf{z}}_{T}|{\bf{z}}_{qT}\right)$ & $\delta\left({\bf{z}}_{T} -  \sum_{q=0}^{Q-1} {\bf{z}}_{qT}\right) $\\
  \hline   $f_{{\bf{z}}_{qT}}$ & $p\left({\bf{z}}_{qT}|{\bf{d}}_{qT}\right)$ & $\delta\left({\bf{z}}_{qT} -  {\bf{b}}_{q} \odot {\bf{d}}_{qT}\right) $\\
  \hline   $f_{{\bf{d}}_{qT}}$ & $p\left({\bf{d}}_{qT}|{\bf{d}}_{qF}\right)$ & $ \delta\left({\bf{d}}_{qT} -  {\bf{F}}^{H}_{MN} {\bf{d}}_{qF}\right) $\\
  \hline   $f_{{\bf{d}}_{qF}}$ & $p\left({\bf{d}}_{qF}|{\bf{c}}_{qF}, {\bf{x}}_{F}\right)$ & $ \delta\left({\bf{d}}_{qF} -  {\bf{x}}_{F} \odot {\bf{c}}_{qF}\right) $\\
  \hline   $f_{{\bf{c}}_{qF}}$ & $p\left({\bf{c}}_{qF}|{\bf{c}}_{q}\right)$ & $ \delta\left({\bf{c}}_{qF}-  {\bf{F}}^{H}_{MN \times L}{\bf{c}}_{q}\right)$\\
  \hline   $f_{{\bf{c}}_{q}}$  & $p\left({\bf{c}}_{q}\right)$ & $N\left({\bf{c}}_{q};0, {\lambda}_{c}{\bf{I}}\right)$\\
  \hline   $f_{{\bf{x}}_{F}}$ & $p\left({\bf{x}}_{F}|{\bf{x}}_{T}\right)$ & $ \delta\left({\bf{x}}_{F}-  {\bf{F}}^{H}_{MN}{\bf{x}}_{T}\right)$\\
  \hline   $f_{{\bf{x}}_{T}}$ & $p\left({\bf{x}}_{T}|{\bf{x}}_{DD}\right)$ & $ \delta\left({\bf{x}}_{T}-  ({\bf{F}}^{H}_N \otimes {\bf{I}}_M){\bf{x}}_{DD}\right)$\\
  \hline   $f_{{\bf{x}}_{DD}}$ & $p\left({\bf{x}}_{DD}|{\bf{x}}_{d}\right)$ & $ \delta\left({\bf{x}}_{DD}-  \sqrt{\rho}{\bf{x}}_p + \sqrt{1 - \rho}{\bf{x}}_d\right)$\\
  \hline   $f_{{\bf{x}}_{d}(i)}$  & $p\left({\bf{x}}_{d}(i)\right)$ & \makecell{$N\left({\bf{x}}_{d}(i); \sum_{\alpha_i \in A}\alpha_i P_i(\alpha_i),  \right.$ \\
  $\left. 1-\left(\sum_{\alpha_i \in A}\alpha_i P_i(\alpha_i)\right)^2 \right)$} \\
 \hline
 \end{tabular}
 \end{center}
\end{table}

The factor graph representation of (\ref{eq:joint_cd}) is shown in Fig. \ref{FIG:RXstructure}, and we will develop message passing algorithms based on the factor graph. In addition, local functions and their corresponding distributions are shown in Table \ref{tbl0}. Note that since there is no prior about ${\bf c}_q$, ${\lambda}_{c} = \infty$. We aim to find the a posteriori distributions (marginals): $p({\bf{c}}_q(i)|\bf{y}_T)$ and $p({\bf{x}}_d(i)|\bf{y}_T)$ and their estimates  in terms of the a posteriori means, i.e., ${\bf \hat c}_q(i) = E({\bf c}_q(i)|{y}_T)$ and ${\bf \hat x}_d(i) = {E}({\bf{x}}_d(i)|{y}_T)$ (based on which hard decisions on the transmitted bits can be made). Throughout this paper, we use $\overleftarrow{\bf{\bar a}}$ and $\overleftarrow{{{\bf{v}}_{{\bf{a}}}}}$ to denote the mean and variance of the Gaussian message about variable $\bf{a}$ in the backward direction, respectively. Similarly, we use $\overrightarrow{\bf{\bar a}}$ and $\overrightarrow{{{\bf{v}}_{{\bf{a}}}}}$ to denote the mean and variance of variable $\bf{a}$ in the forward message passing, respectively.

\subsection{Message Passing Algorithm Design}

For the convenience of message passing algorithm design, we divide the factor graph into four parts as shown in Fig. \ref{FIG:RXstructure}. To make the message computations tractable, mean field \cite{Winn2005} is used in $PART$ $IV$, expected propagation \cite{Minka2001} is used in $PART$ $I$ to deal with the discrete variables ${\bf x}_d$, and the sum-product rule \cite{Loeliger2004} is used in the remaining parts.
In the following, we derive the message computations in these part.

\subsubsection{Message Computations in $PART$ $I$}
We first investigate the backward message computations. As the transmitted symbols are discrete variables, we project the discrete distribution to be Gaussian, i.e.,
\begin{equation}
\overleftarrow{{\bf{\bar x}}_d}\left(i\right)^{'} = \sum_{\alpha_i \in A}{\alpha_i} P_i\left({\alpha_i}\right),
\label{eq:bar_xd1_back}
\end{equation}
\begin{equation}
 \overleftarrow{{{\bf{v}}_{{\bf{x}}_d}}}\left(i\right)^{'} = 1 - | \overleftarrow{{\bf{\bar x}}_d}\left(i\right)|^2,
\label{eq:var_xd1_back}
\end{equation}
where $P_i(\alpha)$ is the a priori probability when ${\bf{x}_d(i)} = \alpha$. In addition, the backward mean $\overleftarrow{{\bf{\bar x}}_d}$ and the backward variance $\overleftarrow{{{\bf{v}}_{{\bf{x}}_d}}}$ can be computed as
\begin{equation}
 \overleftarrow{{{\bf{v}}_{{\bf{x}}_d}}}\left(i\right) = 1/\left(1/\overleftarrow{{{\bf{v}}_{{\bf{x}}_d}}}\left(i\right)^{'} - 1/\overrightarrow{{{\bf{v}}_{{\bf{x}}_d}}}\left(i\right) \right),
\label{eq:var_xd_back}
\end{equation}
\begin{equation}
\overleftarrow{{\bf{\bar x}}_d}\left(i\right) = \overleftarrow{{{\bf{v}}_{{\bf{x}}_d}}}\left(i\right)\left(\overleftarrow{{\bf{\bar x}}_d}\left(i\right)^{'} /\overleftarrow{{{\bf{v}}_{{\bf{x}}_d}}}\left(i\right)^{'} - \overrightarrow{{{{\bf{\bar x}}_d}}}\left(i\right)/\overrightarrow{{{\bf{v}}_{{\bf{x}}_d}}}\left(i\right)\right),
\label{eq:bar_xd_back}
\end{equation}
where $\overrightarrow{{{{\bf{\bar x}}_d}}}$ and $\overrightarrow{{{\bf{v}}_{{\bf{x}}_d}}}$ are computed by (\ref{eq:bar_xd1_for}) and (\ref{eq:var_xd1_for}) in the last iteration.
With (\ref{eq:xDD}), we have
\begin{equation}
 \overleftarrow{{\bf{\bar x}}_{DD}} = \sqrt{\rho}{\bf{x}}_p + \sqrt{1 - \rho} \overleftarrow{{\bf{\bar x}}_d},
\label{eq:bar_xDD_back}
\end{equation}
\begin{equation}
 \overleftarrow{{{\bf{v}}_{{\bf{x}}_{DD}}}} = \left(1 - \rho\right)\overleftarrow{{{\bf{v}}_{{\bf{x}}_d}}}.
\label{eq:var_xDD_back}
\end{equation}
It can be seen from (\ref{eq:x_T}) that ${\bf{x}}_T $ is obtained by the Fourier transform of ${\bf{X}}_{DD}$. Thus, we have
\begin{equation}
\overleftarrow{{\bf{\bar X}}_{DD}} = {\bf{vec}}^{-1}\left(\overleftarrow{{\bf{\bar x}}_{DD}}\right),
\label{eq:bar_XDD_back}
\end{equation}
\begin{equation}
 \overleftarrow{{\bf{\bar x}}_{T}} = {\bf{vec}}\left(\overleftarrow{{\bf{\bar X}}_{DD}}{{\bf{F}}^H_N}\right).
\label{eq:bar_xT_back}
\end{equation}
We make approximation to the computation of variance by averaging the variances of ${\bf{x}}_{T}$, i.e.,
\begin{equation}
 \overleftarrow{{{v}_{{\bf{x}}_{T}}}} = <\overleftarrow{{{\bf{v}}_{{\bf{x}}_{DD}}}}>.
\label{eq:var_xT_back}
\end{equation}
When using (\ref{eq:var_xT_back}) for approximation, the matrix operation is avoided, significantly reducing computational complexity.
According to (\ref{eq:x_F}), we can get
\begin{equation}
 \overleftarrow{{\bf{\bar x}}_{F}} = {\bf{F}}_{MN}  \overleftarrow{{\bf{\bar x}}_{T}},
\label{eq:bar_xTF_back}
\end{equation}
\begin{equation}
 \overleftarrow{{{v}_{{\bf{x}}_{F}}}} =  \overleftarrow{{{v}_{{\bf{x}}_{T}}}}.
\label{eq:var_xTF_back}
\end{equation}

Next, we study the forward message computations. With belief propagation and Fig. \ref{FIG:RXstructure}, the forward mean $\overrightarrow{{\bf{\bar x}}_{F}}$ and forward variance $\overrightarrow{{{\bf{v}}_{{\bf{x}}_{F}}}}$ can be computed as
\begin{equation}
 \overrightarrow{{{\bf{v}}_{{\bf{x}}_{F}}}} =  1./\left(\sum_{q=0}^{Q-1} 1./ \overrightarrow{{{\bf{v}}_{{\bf{x}}_{qF}}}}\right),
\label{eq:var_xTF_for}
\end{equation}
\begin{equation}
 \overrightarrow{{\bf{\bar x}}_{F}} =  \overrightarrow{{\bf{v}_{{\bf{x}}_{F}}}} \odot \left(\sum_{q=0}^{Q-1} \overrightarrow{{\bf{\bar x}}_{qF}} ./ \overrightarrow{{{\bf{v}}_{{\bf{x}}_{qF}}}}\right).
\label{eq:bar_xTF_for}
\end{equation}
Similar to the backward message computations, we can get
\begin{equation}
 \overrightarrow{{\bf{\bar x}}_{T}} = {\bf{F}}_{MN}^{H} \overrightarrow{{\bf{\bar x}}_{F}},
\label{eq:bar_xT_for}
\end{equation}
\begin{equation}
 \overrightarrow{{{v}_{{\bf{x}}_{T}}}} = < \overrightarrow{{{\bf{v}}_{{\bf{x}}_{F}}}}>.
\label{eq:var_xT_for}
\end{equation}
\begin{equation}
 \overrightarrow{{\bf{\bar X}}_{T}} = {\bf{vec}}^{-1}\left(\overrightarrow{{\bf{\bar x}}_{T}}\right),
\label{eq:bar_XT_for}
\end{equation}
\begin{equation}
\overrightarrow{{\bf{\bar x}}_{DD}} = {\bf{vec}}\left(\overrightarrow{{\bf{\bar X}}_{T}}{{\bf{F}}_N}\right),
\label{eq:bar_xDD_for}
\end{equation}
\begin{equation}
 \overrightarrow{{{v}_{{\bf{x}}_{DD}}}} =  \overrightarrow{{{v}_{{\bf{x}}_{T}}}}.
\label{eq:var_xDD_for}
\end{equation}
\begin{equation}
 \overrightarrow{{\bf{\bar x}}_{d}} = \left(\overrightarrow{{\bf{\bar x}}_{DD}} -  \sqrt{\rho}{\bf{x}}_p\right)/\sqrt{1 - \rho} ,
\label{eq:bar_xd1_for}
\end{equation}
\begin{equation}
 \overrightarrow{{{\bf{v}}_{{{x}}_{d}}}} = \overrightarrow{{{v}_{{\bf{x}}_{DD}}}}/\left(1 - \rho\right).
\label{eq:var_xd1_for}
\end{equation}
After obtaining the forward mean $\overrightarrow{{\bf{\bar x}}_{d}}$ and forward variance $\overrightarrow{{{{v}}_{{\bf{x}}_{d}}}}$, the probability of ${\bf{x}}_d$ can be updated as
\begin{equation}
P_i\left({\alpha_i}\right) \propto \exp \left\{- \frac{|\alpha_i -  \overrightarrow{{\bf{\bar x}}_{d}}(i)|^2}{ \overrightarrow{{{{v}}_{{\bf{x}}_{d}}}}}\right\} .
\label{eq:Pm_alpha_for}
\end{equation}
Moreover, we can obtain the estimate of ${\bf{x}}_d$ through
\begin{equation}
{\bf{\hat x}}_d\left(i\right) = argmax_{\alpha_i}\{P_i(\alpha_i)\}.
\label{eq:hat_xd}
\end{equation}

\subsubsection{Message Computations in $PART$ $II$}
We first look at the backward message computations. According to (\ref{eq:d_T}), the backward mean $\overleftarrow{{\bf{\bar d}}_{qT}}$ and backward variance $\overleftarrow{{{{v}}_{{\bf{d}}_{qT}}}}$ can be computed as
\begin{equation}
 \overleftarrow{{\bf{\bar d}}_{qT}} =  {\bf{F}}^{H}_{MN}\overleftarrow{{\bf{\bar d}}_{qF}} ,
\label{eq:bar_dqT_back}
\end{equation}
\begin{equation}
 \overleftarrow{{{{v}}_{{\bf{d}}_{qT}}}} =  < \overleftarrow{{{\bf{v}}_{{\bf{d}}_{qF}}}}>.
\label{eq:var_dqT_back}
\end{equation}
Considering (\ref{eq:z_qT}) and constant vector ${\bf{b}}_{q}$, we can get
\begin{equation}
 \overleftarrow{{\bf{\bar z}}_{qT}}^{'} =  \overleftarrow{{\bf{\bar d}}_{qT}} \odot {\bf{b}}_{q},
\label{eq:bar_zqT1_back}
\end{equation}
\begin{equation}
 \overleftarrow{{{{v}}_{{\bf{z}}_{qT}}}}^{'} =   <\overleftarrow{{{{v}}_{{\bf{d}}_{qT}}}}\left({\bf{b}}_{q} \odot {\bf{b}}_{q}^{*}\right)>.
\label{eq:var_zqT1_back}
\end{equation}
Moreover, we use the damping to improve the robustness of the algorithm, i.e.,
\begin{equation}
 \overleftarrow{{{{v}}_{{\bf{z}}_{qT}}}}= 1./\left(\left(1-\eta\right)/\overleftarrow{{{{v}}_{{\bf{z}}_{qT}}}}_{pre} + \eta / \overleftarrow{{{{v}}_{{\bf{z}}_{qT}}}}^{'}\right),
\label{eq:var_zqT_back}
\end{equation}
\begin{equation}
 \overleftarrow{{\bf{\bar z}}_{qT}} =  \overleftarrow{{{{v}}_{{\bf{z}}_{qT}}}}\left(\left(1-\eta\right)\overleftarrow{{\bf{\bar z}}_{qT}}_{pre}/\overleftarrow{{{{v}}_{{\bf{z}}_{qT}}}}_{pre} + \eta \overleftarrow{{\bf{\bar z}}_{qT}}^{'}/ \overleftarrow{{{{v}}_{{\bf{z}}_{qT}}}}^{'}\right) ,
\label{eq:bar_zqT_back}
\end{equation}
\begin{equation}
 \overleftarrow{{{{v}}_{{\bf{z}}_{qT}}}}_{pre} = \overleftarrow{{{{v}}_{{\bf{z}}_{qT}}}},
\label{eq:var_zqT_back_pre}
\end{equation}
\begin{equation}
 \overleftarrow{{\bf{\bar z}}_{qT}}_{pre} =   \overleftarrow{{\bf{\bar z}}_{qT}},
\label{eq:bar_zqT_back_pre}
\end{equation}
where $0 \leq \eta \leq 1$ is the damping factor.
According to the update rule for the sum operation \cite{Loeliger2004}, the backward mean $\overleftarrow{{\bf{\bar z}}_{T}}$ and backward variance $\overleftarrow{{{\bf{v}}_{{\bf{z}}_{T}}}}$ can be computed as
\begin{equation}
 \overleftarrow{{\bf{\bar z}}_{T}} = \sum_{q = 0}^{Q-1}\overleftarrow{{\bf{\bar z}}_{qT}},
\label{eq:bar_zT_back}
\end{equation}
\begin{equation}
 \overleftarrow{{{\bf{v}}_{{\bf{z}}_{T}}}} = \sum_{q = 0}^{Q-1} \overleftarrow{{{\bf{v}}_{{{z}}_{qT}}}}.
\label{eq:var_zT_back}
\end{equation}

We then study the forward message computations. It is clear that
\begin{equation}
 \overrightarrow{{\bf{\bar z}}_{T}} = {{\bf{y}}_{T}},
\label{eq:bar_zT_for}
\end{equation}
\begin{equation}
 \overrightarrow{{{{v}}_{{\bf{z}}_{T}}}} = w^{-1}.
\label{eq:var_zT_for}
\end{equation}
According to the update rule in \cite{Loeliger2004}, we have
\begin{equation}
 \overrightarrow{{\bf{\bar z}}_{qT}} = \overrightarrow{{\bf{\bar z}}_{T}} -  \overleftarrow{{\bf{\bar z}}_{T}} +  \overleftarrow{{\bf{\bar z}}_{qT}},
\label{eq:bar_zqT_for}
\end{equation}
\begin{equation}
 \overrightarrow{{{{v}}_{{\bf{z}}_{qT}}}} = \overrightarrow{{{{v}}_{{\bf{z}}_{T}}}} +  \overleftarrow{{{{v}}_{{\bf{z}}_{T}}}} -  \overleftarrow{{{{v}}_{{\bf{z}}_{qT}}}}.
\label{eq:var_zqT_for}
\end{equation}
Similar to (\ref{eq:bar_dqT_back})-(\ref{eq:var_zqT1_back}), we can obtain
\begin{equation}
 \overrightarrow{{\bf{\bar d}}_{qT}} = \overrightarrow{{\bf{\bar z}}_{qT}}./{\bf{b}}_{q},
\label{eq:bar_dqT_for}
\end{equation}
\begin{equation}
 \overrightarrow{{{\bf{v}}_{{{d}}_{qT}}}} =  \overrightarrow{{{{v}}_{{\bf{z}}_{qT}}}}./\left({\bf{b}}_{q} \odot {\bf{b}}_{q}^{*} \right),
\label{eq:var_dqT_for}
\end{equation}
\begin{equation}
 \overrightarrow{{\bf{\bar d}}_{qF}} =  {\bf{F}}_{MN}\overrightarrow{{\bf{\bar d}}_{qT}} ,
\label{eq:bar_dqF_for}
\end{equation}
\begin{equation}
 \overrightarrow{{{{v}}_{{{d}}_{qF}}}} =  < \overrightarrow{{{\bf{v}}_{{{d}}_{qT}}}}>.
\label{eq:var_dqF_for}
\end{equation}

\subsubsection{Message Computations in $PART$ $III$}
Again, for the forward message computations, we define the intermediate vector ${\bf{c}}_{qL} = [{\bf{c}}^T_q, {\bf{0}}^T]^T$. Thus, we have
\begin{equation}
{\bf{c}}_{qF} = {\bf{F}}_{MN \times L} {\bf{c}}_{q} =  {\bf{F}}_{MN} {\bf{c}}_{qL},
\label{eq:cqF2cqL_ex}
\end{equation}
and it is clear that
\begin{equation}
\overrightarrow{{\bf{\bar c}}_{qF}} = {\bf{F}}_{MN}[\overrightarrow{{\bf{\bar c}}_{q}}, {\bf{0}}^T]^T,
\label{eq:bar_cqF_for}
\end{equation}
\begin{equation}
 \overrightarrow{{{\bf{v}}_{{\bf{c}}_{qF}}}}  = \left(L/MN\right)\overrightarrow{{{\bf{v}}_{{\bf{c}}_{q}}}} .
\label{eq:var_cqF_for}
\end{equation}
Then we investigate the backward message computations. Considering (\ref{eq:cqF2cqL_ex}) and ${\bf{c}}_{qL}$, we have
\begin{equation}
\overleftarrow{{\bf{\bar c}}_{qL}} = {\bf{F}}^{H}_{MN}\overleftarrow{{\bf{\bar c}}_{qF}},
\label{eq:bar_cq1_back}
\end{equation}
\begin{equation}
\overleftarrow{{\bf{\bar c}}_{q}} = \overleftarrow{{\bf{\bar c}}_{qL}}\left(1 : L\right),
\label{eq:bar_cq_back}
\end{equation}
\begin{equation}
 \overleftarrow{{{{v}}_{{\bf{c}}_{q}}}}  = < \overleftarrow{{{\bf{v}}_{{\bf{c}}_{qF}}}} > .
\label{eq:var_cq_back}
\end{equation}
In addition, $f_{{\bf{c}}_{q}} = N({\bf{c}}_{q};0, {\lambda}_{c}{\bf{I}})$ and $\lambda = \infty $. Thus, we obtain
\begin{equation}
{{\bf{\hat c}}_{q}} =  \overleftarrow{{\bf{\bar c}}_{q}},
\label{eq:bar_cq_for}
\end{equation}
\begin{equation}
{{{\bf{v}}_{{\bf{\hat c}}_{q}}}} = \overleftarrow{{{\bf{v}}_{{\bf{c}}_{qF}}}} .
\label{eq:var_cq_for}
\end{equation}

\subsubsection{Message Computations in $PART$ $IV$}
With belief propagation and Fig. \ref{FIG:RXstructure}, the backward mean $ \overleftarrow{{\bf{\bar x}}_{qF}}$ and backward variance $ \overleftarrow{{{{v}}_{{\bf{x}}_{qF}}}}$ can be computed as
\begin{equation}
 \overleftarrow{{{{v}}_{{\bf{x}}_{qF}}}}  = 1/\left(1/\overleftarrow{{{{v}}_{{\bf{x}}_{F}}}} + \sum_{j = 0, j \neq q}^{Q-1} 1/\overrightarrow{{{{v}}_{{\bf{x}}_{qF}}}}\right),
\label{eq:var_xqTF_back}
\end{equation}
\begin{equation}
 \overleftarrow{{\bf{\bar x}}_{qF}} = \overleftarrow{{{{v}}_{{\bf{x}}_{qF}}}}\left(\overleftarrow{{\bf{\bar x}}_{F}} /\overleftarrow{{{{v}}_{{\bf{x}}_{F}}}} + \sum_{j = 0, j \neq q}^{Q-1} \overrightarrow{{\bf{\bar x}}_{qF}} /\overrightarrow{{{{v}}_{{\bf{x}}_{qF}}}}\right).
\label{eq:bar_xqTF_back}
\end{equation}
The approximate a posteriori mean ${{\bf{\hat x}}_{qF}}$ and variance ${{{{v}}_{{\bf{\hat x}}_{qF}}}}$ can be obtained as
\begin{equation}
 {{{{v}}_{{\bf{\hat x}}_{qF}}}}  = 1/\left(1/\overleftarrow{{{{v}}_{{\bf{x}}_{qTF}}}}  + 1/\overrightarrow{{{{v}}_{{\bf{x}}_{qF}}}}\right),
\label{eq:vhat_xqTF_for}
\end{equation}
\begin{equation}
 {{\bf{\hat x}}_{qF}} = {{{{v}}_{{\bf{\hat x}}_{qF}}}}\left(\overleftarrow{{\bf{\bar x}}_{qF}}/\overleftarrow{{{{v}}_{{\bf{x}}_{qF}}}}  + \overrightarrow{{\bf{\bar x}}_{qF}}/\overrightarrow{{{{v}}_{{\bf{x}}_{qF}}}}\right).
\label{eq:hat_xqTF_for}
\end{equation}
After updating ${{\bf{\hat x}}_{qF}}$ and ${{{{v}}_{{\bf{\hat x}}_{qF}}}}$, the backward mean $\overleftarrow{{\bf{\bar c}}_{qF}}$ and backward variance $\overleftarrow{{{\bf{v}}_{{\bf{c}}_{qF}}}}$ can be obtained by the mean field rule \cite{Riegler2013, Guo2017Low} as
\begin{equation}
\overleftarrow{{\bf{\bar c}}_{qF}}(i) = \frac{\overrightarrow{{\bf{\bar d}}_{qF}}(i){{\bf{\hat x}}_{qTF}}^{*}(i)}{|{{\bf{\hat x}}_{qTF}}(i)|^2 +  {{{{v}}_{{\bf{\hat x}}_{qTF}}}}},
\label{eq:bar_cqF_back}
\end{equation}
\begin{equation}
\overleftarrow{{{\bf{v}}_{{\bf{c}}_{qF}}}}(i)= \frac{\overrightarrow{{{{v}}_{{{d}}_{qF}}}}}{|{{\bf{\hat x}}_{qTF}}(i)|^2 +  {{{{v}}_{{\bf{\hat x}}_{qTF}}}}}.
\label{eq:var_cqF_back}
\end{equation}
Similarly, we can compute the forward mean $\overrightarrow{{\bf{\bar x}}_{qF}}$ and forward variance $\overrightarrow{{{\bf{v}}_{{\bf{x}}_{qF}}}}$ through the mean field rule, i.e.,
\begin{equation}
 {{{{v}}_{{\bf{\hat c}}_{qF}}}}  = 1/\left(1/\overleftarrow{{{{v}}_{{\bf{c}}_{qF}}}}  + 1/\overrightarrow{{{{v}}_{{\bf{c}}_{qF}}}}\right),
\label{eq:vhat_cqF_for}
\end{equation}
\begin{equation}
 {{\bf{\hat c}}_{qF}} = {{{{v}}_{{\bf{\hat c}}_{qF}}}}\left(\overleftarrow{{\bf{\bar c}}_{qF}}/\overleftarrow{{{{v}}_{{\bf{c}}_{qF}}}}  + \overrightarrow{{\bf{\bar c}}_{qF}}/\overrightarrow{{{{v}}_{{\bf{c}}_{qF}}}}\right),
\label{eq:hat_cqF_for}
\end{equation}
\begin{equation}
\overrightarrow{{\bf{\bar x}}_{qF}}(i) = \frac{\overrightarrow{{\bf{\bar d}}_{qF}}(i){{\bf{\hat c}}_{qF}}^{*}(i)}{|{{\bf{\hat c}}_{qF}}(i)|^2 +  {{{{v}}_{{\bf{\hat c}}_{qF}}}}},
\label{eq:bar_xqTF_for}
\end{equation}
\begin{equation}
\overrightarrow{{{\bf{v}}_{{\bf{x}}_{qF}}}}(i)= \frac{\overrightarrow{{{{v}}_{{{d}}_{qF}}}}}{|{{\bf{\hat c}}_{qF}}(i)|^2 +  {{{{v}}_{{\bf{\hat c}}_{qF}}}}}.
\label{eq:var_xqTF_for}
\end{equation}
Then, the backward message $\overleftarrow{{\bf{\bar d}}_{qF}}$ and $\overleftarrow{{{\bf{v}}_{{\bf{d}}_{qF}}}}$ can be computed as
\begin{equation}
\overleftarrow{{\bf{\bar d}}_{qF}} = \overrightarrow{{\bf{\bar c}}_{qF}} \odot \overleftarrow{{\bf{\bar x}}_{qF}},
\label{eq:bar_dqF_back}
\end{equation}
\begin{equation}
\overleftarrow{{{\bf{v}}_{{\bf{d}}_{qF}}}} =  \overleftarrow{{{{v}}_{{\bf{x}}_{qF}}}}\left(\overrightarrow{{\bf{\bar c}}_{qF}} \odot \overrightarrow{{\bf{\bar c}}_{qF}}^{*}\right) + \overrightarrow{{{{v}}_{{\bf{c}}_{qF}}}}\left(\overleftarrow{{\bf{\bar x}}_{qF}} \odot \overleftarrow{{\bf{\bar x}}_{qF}}^{*}\right) + \overrightarrow{{{{v}}_{{\bf{c}}_{qF}}}}\overleftarrow{{{{v}}_{{\bf{x}}_{qF}}}}{\bf{1}}.
\label{eq:var_dqF_back}
\end{equation}
The main steps of the algorithm for the SP-DD receiver are summarized in Algorithm \ref{alg1}.

\begin{algorithm}[htpb]
	\caption{Algorithm for the SP-DD Receiver}
	\renewcommand{\algorithmicrequire}{\textbf{Input:}}
	\renewcommand{\algorithmicensure}{\textbf{Output:}}
	
	\label{alg1}
	\begin{algorithmic}[1]
    \REQUIRE{${\bf y}_T$, ${\bf x}_p$, $w^{-1}$}
    \ENSURE {${\bf{\hat x}}_p$, ${{\bf{\hat c}}_{q}}$}
    \STATE Initialization: $P_m = {\bf{0}}$, $\eta = 0.8$, $\overrightarrow{{\bf{\bar x}}_{qF}} = \bf{0}$, $\overrightarrow{{{{v}}_{{\bf{x}}_{qF}}}} = \infty$, $\overleftarrow{{\bf{\bar z}}_{T}} = \bf{0}$, $\overleftarrow{{{{v}}_{{\bf{z}}_{T}}}} = \bf{0}$, $\overleftarrow{{\bf{\bar z}}_{qT}} = \bf{0}$, $\overleftarrow{{{{v}}_{{\bf{z}}_{qT}}}} = 0$, $\overrightarrow{{\bf{\bar z}}_{T}} = {\bf y}_T$, $\overrightarrow{{{{v}}_{{\bf{z}}_{qT}}}} = w^{-1}$, $\overrightarrow{{\bf{\bar c}}_{qF}} = \bf{0}$,  $\overrightarrow{{{\bf{v}}_{{\bf{c}}_{qF}}}} = \infty \bf{I}$.
    \REPEAT
    \STATE Compute  $\overleftarrow{{\bf{\bar x}}_{F}}$ and $\overleftarrow{{{v}_{{\bf{x}}_{F}}}}$ through (\ref{eq:bar_xd1_back}) to (\ref{eq:var_xTF_back})
    \STATE Obtain  $\overrightarrow{{\bf{\bar d}}_{qF}} $ and $\overrightarrow{{{{v}}_{{{d}}_{qF}}}}$ through (\ref{eq:bar_zT_for}) to (\ref{eq:var_dqF_for})
    \STATE Calculate  $\overleftarrow{{\bf{\bar x}}_{qF}}$ and $\overleftarrow{{{{v}}_{{\bf{x}}_{qF}}}}$ as in (\ref{eq:var_xqTF_back}) and (\ref{eq:bar_xqTF_back})
    \STATE Update  ${{\bf{\hat x}}_{qF}}$ and ${{{{v}}_{{\bf{\hat x}}_{qF}}}}$ through (\ref{eq:vhat_xqTF_for}) and (\ref{eq:hat_xqTF_for})
    \STATE Compute   $\overleftarrow{{\bf{\bar c}}_{qF}}$ and $\overleftarrow{{{\bf{v}}_{{\bf{c}}_{qF}}}}$  using  (\ref{eq:bar_cqF_back}) to (\ref{eq:var_cqF_back})
    \STATE Update  ${{\bf{\hat c}}_{qF}}$ and ${{{{v}}_{{\bf{\hat c}}_{qF}}}}$ as in  (\ref{eq:vhat_cqF_for}) and (\ref{eq:hat_cqF_for})
    \STATE Calculate $\overrightarrow{{\bf{\bar x}}_{qF}}$ and $\overrightarrow{{{\bf{v}}_{{\bf{x}}_{qF}}}}$ using  (\ref{eq:bar_xqTF_for}) and (\ref{eq:var_xqTF_for})
    \STATE Compute $\overrightarrow{{\bf{\bar c}}_{qF}}$ and $\overrightarrow{{{\bf{v}}_{{\bf{c}}_{qF}}}}$ using  (\ref{eq:bar_cqF_for}) and (\ref{eq:var_cqF_for})
    \STATE Obtain ${{\bf{\hat c}}_{q}}$ and ${{{\bf{v}}_{{\bf{\hat c}}_{q}}}}$ using (\ref{eq:bar_cq1_back}) to (\ref{eq:var_cq_for})
    \STATE Compute $\overleftarrow{{\bf{\bar d}}_{qF}}$ and $\overleftarrow{{{\bf{v}}_{{\bf{d}}_{qF}}}}$ as in  (\ref{eq:bar_dqF_back}) and (\ref{eq:var_dqF_back})
    \STATE Update $\overleftarrow{{\bf{\bar z}}_{qT}}$ and  $\overleftarrow{{{{v}}_{{\bf{z}}_{qT}}}}$ through (\ref{eq:bar_dqT_back}) to (\ref{eq:bar_zqT_back})
    \STATE Calculate $\overleftarrow{{\bf{\bar z}}_{T}}$ and  $\overleftarrow{{{\bf{v}}_{{\bf{z}}_{T}}}}$ using (\ref{eq:bar_zT_back}) and (\ref{eq:var_zT_back})
    \STATE Update $P_i$ through (\ref{eq:var_xTF_for}) to (\ref{eq:Pm_alpha_for})
    \STATE Obtain the ${{\bf{\hat x}}_{d}}$ as in (\ref{eq:hat_xd})
    \UNTIL{terminated}
	\end{algorithmic}
\end{algorithm}

\subsection{Iterative Receiver with Designed Pilots}\label{SP-FD}

According to $PART$ $IV$, ${\bf x}_{F}$ and ${\bf d}_{qF}$ are symbols in the frequency domain, so the CE of the proposed SP-DD receiver is carried out in the frequency domain.
Moreover, the initial CE is only based on pilot symbols as the estimates of data symbols are not available. Therefore, the pilot symbol power in the frequency domain can affect the accuracy of the initial CE.
Therefore, we design periodic pilots in the time domain to achieve power concentration in the frequency domain and propose an SP-DD-D receiver to improve the accuracy of the initial CE.

For the convenience of description, we define the following auxiliary variables. The DD domain data symbols are denoted as ${\bf{X}}_{DD;d} = {\bf{vec}^{-1}}({\bf{x}}_{d})$, and the time domain data signal is represented as ${\bf{x}}_{T;d} = {\bf{vec}}({\bf{X}}_{DD;d}{{\bf{F}}^H_N})$. After the DFT, ${\bf{x}}_{F;d}$ is given as
\begin{equation}
{\bf{x}}_{F;d} = {\bf{F}}_{MN}{\bf{x}}_{T;d}.
\label{eq:bar_x_TFd}
\end{equation}
Similar to the data symbols, the DD domain pilots can be express as ${\bf{X}}_{DD;p} = {\bf{vec}^{-1}}({\bf{x}}_{p})$, and the time domain signal can be expressed as ${\bf{x}}_{T;p} = {\bf{vec}}({\bf{X}}_{DD;p}{{\bf{F}}^H_N})$. After the DFT, ${\bf{x}}_{F;p}$ is given as
\begin{equation}
{\bf{x}}_{F;p} = {\bf{F}}_{MN}{\bf{x}}_{T;p}.
\label{eq:bar_x_TFp}
\end{equation}
Thus, we have
\begin{equation}
{\bf{x}}_{F} = \sqrt{{{\bf 1}}-{{\bm \rho}_F}} \odot {\bf{x}}_{F;d} + \sqrt{{\bm \rho}_F} \odot {\bf{x}}_{F;p},
\label{eq:bar_x_TF_all}
\end{equation}
where ${\bm {\rho}}_F$ is a power allocation factor vector.
We aim to concentrate the power of pilot symbols in the frequency domain. To achieve this, we first design the frequency domain pilot symbols and then transform them to obtain the DD domain pilot symbols. We set $P = MN/{\beta}$, where $P$ and $\beta$ are integers, and $\beta$ is a power concentration factor, i.e., the power of each $\beta$ pilot is concentrated to be one.

In addition, $P > L$ is required to facilitate the CE. Hence, we have
\begin{equation}
{{\bf{x}}_{F;p}(i)} = \left\{ {\begin{array}{*{10}{c}}
{ {\bf{x}}_{F;sp}(i),\mbox{if}\ mod(i, \beta) = 0}\\
{ 0 , \mbox{if}\ mod(i, \beta) \neq 0}
\end{array}} \right. .
\label{eq:bf_xp_FD}
\end{equation}
where ${\bf{x}}_{F; sp}$ represents a non-zero pilot sequence in the frequency domain with length $P$.

\begin{figure}[htbp]
	\centering
        {\includegraphics[scale=.06]{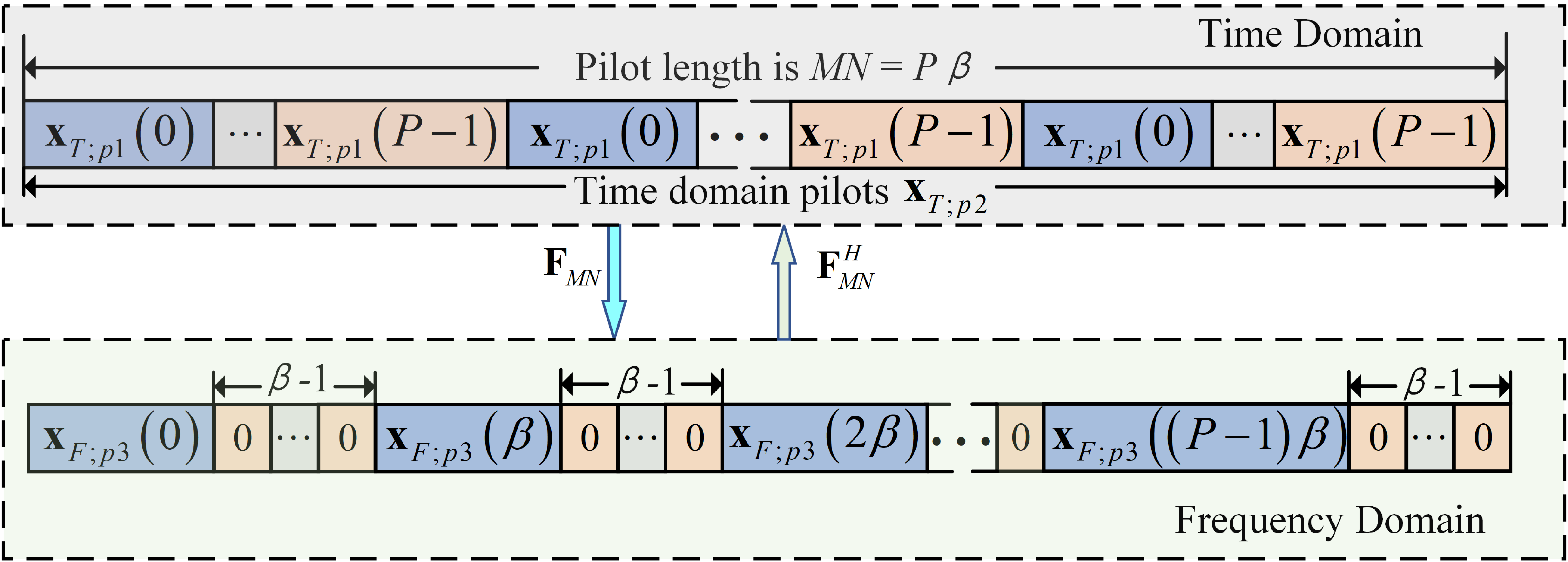}}
	\caption{Schematic diagram of time-domain pilot and frequency-domain pilot conversion
	\label{FIG:PT2F}}
\end{figure}

According to the properties of the DFT, we can achieve the frequency domain sequence in (\ref{eq:bf_xp_FD}) by generating a periodic sequence in the time domain. Firstly, we generate an constant power sequence ${\bf{x}}_{T;p1}$ with a length of $P$ and periodically extend the sequence to obtain ${\bf{x}}_{T;p2}$ with a length of $MN$, i.e.,
\begin{equation}
 {\bf{x}}_{T;p2}(i) = {\bf{x}}_{T;p1}(i),
\label{eq:xp_Plen1}
\end{equation}
where $i \in [0, P-1]$,
\begin{equation}
 {\bf{x}}_{T;p2}\left(i + mP\right) = {\bf{x}}_{T;p2}(i),
\label{eq:xp_Plen}
\end{equation}
and $m \in [0, \beta-1]$. Next, we obtain ${\bf{x}}_{F;p3}$ using the DFT operation on ${\bf{x}}_{T;p2}$, i.e.,
\begin{equation}
 {\bf{x}}_{F;p3} = {\bf{F}}_{MN}{\bf{x}}_{T;p2}.
\label{eq:xp_Plen_F}
\end{equation}
It is noted that
\begin{equation}
 {\bf{x}}_{F;p3}(i) =  \begin{array}{*{10}{c}}
{ 0 , \mbox{if}\ mod(i, \beta) \neq 0}
\end{array},
\label{eq:xp_Plen_F1}
\end{equation}
and
\begin{equation}
\begin{aligned}
{\bf{x}}_{F;p3}(i\beta) & = \sqrt{\frac{1}{P\beta}}\sum_{n = 0}^{P\beta - 1}{\bf{x}}_{p2}(n)exp\left(-\frac{j2\pi in}{P}\right)\\
&= \sqrt{\beta}\sqrt{\frac{1}{P}}\sum_{n = 0}^{P\beta - 1}{\bf{x}}_{p2}(n)exp\left(-\frac{j2\pi in}{P}\right).
\label{eq:xp_Plen_F2}
\end{aligned}
\end{equation}
Finally, we use ${\bf{x}}_{p3}$ as the pilot sequence in the frequency domain, i.e,
\begin{equation}
{\bf{x}}_{F;p} =  {\bf{x}}_{F;p3}.
\label{eq:xp_F_Fin}
\end{equation}
The relationship between ${\bf{x}}_{T;p1}$, ${\bf{x}}_{T;p2}$ and ${\bf{x}}_{F;p3}$ is shown in Fig. \ref{FIG:PT2F}.
 Considering (\ref{eq:bar_x_TF_all}), (\ref{eq:xp_Plen_F1}), and (\ref{eq:xp_F_Fin}), the the power allocation factor in the frequency domain can be expressed as
\begin{equation}
{{{\bf {\rho}}_F}( i )} = \left\{ {\begin{array}{*{10}{c}}
{ \rho_F,\mbox{if}\ mod(i, \beta) = 0}\\
{ 0 , \mbox{if}\ mod(i, \beta) \neq 0}
\end{array}} \right. .
\label{eq:bf_rho}
\end{equation}
With (\ref{eq:xp_Plen_F2}), Fig .\ref{FIG:PT2F} and the Parseval's theorem, the power of non-zero elements in ${\bf{x}}_{F;p3}$ is $\beta$ times that of elements in ${\bf{x}}_{T;p2}$, i.e., ${{\rho}}_F = \beta \rho $. Thus, through (\ref{eq:xp_Plen_F}), the concentration of pilot power in the frequency domain is achieved.

After obtaining the time domain pilots through (\ref{eq:xp_Plen1}) and (\ref{eq:xp_Plen}), we can acquire the pilots in the DD domain as
\begin{equation}
{\bf{X}}_{T;p2} = {\bf{vec}}^{-1}\left({\bf{x}}_{T;p2}\right),
\label{eq:xp_T2}
\end{equation}
\begin{equation}
{\bf{x}}_{DD;p} = {\bf{vec}}\left({\bf{X}}_{T;p2}{{\bf{F}}_N}\right).
\label{eq:xp_DD_D}
\end{equation}

Moreover, in the initial CE, we only calculate the value of $\overleftarrow{{\bf{\bar c}}_{qF}}$ when the pilot is non-zero. In addition, as the pilot is generated in the time domain, the PAPR of the SP-DD-D receiver will be smaller compared to the PAPR of the SP-DD receiver, especially when the $\rho$ value is large. The main difference lies in the pilot generation method. Table \ref{tbld2r} summarizes the differences between the two receivers.

\begin{table}[htbp]
 \caption{\label{tbld2r} Difference between the SP-DD receiver and the SP-DD-D receiver}
 \begin{center}
 \begin{tabular}{c|c|c}
  \hline \textbf{Item} & \textbf{SP-DD receiver} & \textbf{SP-DD-D receiver} \\
  \hline \makecell{Pilot generation} & \makecell{DD domain} & \makecell{Time domain} \\
  \hline \makecell{Pilot distribution \\in frequency domain}& \makecell{All subcarriers} & Part of subcarriers\\
  \hline \makecell{Changes in PAPR}  &\makecell{--} & \makecell{Smaller than SP-DD } \\
  \hline \makecell{Initial CE accuracy}  &\makecell{--} & \makecell{Better than SP-DD} \\
 \hline
 \end{tabular}
 \end{center}
\end{table}

\section{Pilot Power Ratio Determination and Complexity Analysis}\label{PA}
In this section, we analyze the complexity of the proposed method and determine pilot power factor. We note that, both the SP-DD receiver and the SP-DD-D receiver are iterative algorithms, and accurate performance analysis is very challenging.
To enable the analysis, we make the following assumptions: (1) for each branch, ${\bf{\hat z}}_{qT}$ is approximately ${\bf{z}}_{qT}$ (without error) and (2) $\sum_{q = 0}^{Q-1}\kappa_q = 1$, where $\kappa_q$ represents the average power of the $q$-th branch data.

From (\ref{eq:d_T}) and (\ref{eq:z_qT}), we have
\begin{equation}
{\bf{d}}_{qF}^{init} = {\bf{F}}_{MN}diag\{{\bf{b}}_q\}^{-1}{\bf{z}}_{qT}.
\label{eq:d_qF_init}
\end{equation}
Define ${\bf{d}}_{qF;k}^{init}(i) = {\bf{d}}_{qF}^{init}(i\beta + k)$ and ${\bf{x}}_{F;d;k}^{init}(i) = {\bf{x}}_{qF}^{init}(k\beta + i)$. According to (\ref{eq:d_qF}) and (\ref{eq:bar_x_TF_all}), ${\bf{d}}_{qF;i}^{init}$ can be rewritten as,
\begin{equation}
\begin{aligned}
{\bf{d}}_{qF;0}^{init} &= \sqrt{\frac{\rho_F}{\beta}}diag\{{\bf{x}}_{F;sp}\}{\bf F}_{P\times L}{\bf c}_{qL} \\
&+ \sqrt{\frac{1 - \rho_F}{\beta}}diag\{{\bf{x}}_{F;d;0}\}{\bf F}_{P\times L}{\bf c}_{qL} + {\bf w}_{q;0},
\label{eq:x_F_init1_0}
\end{aligned}
\end{equation}
\begin{equation}
\begin{aligned}
{\bf{d}}_{qF;k}^{init} = \sqrt{\frac{1}{\beta}}diag\{{\bf x}_{F;d;k}\}{\bf F}_{P\times L}\Lambda_i{\bf c}_{qL} + {\bf w}_{q;k}, k \neq 0,
\label{eq:x_F_init1_i}
\end{aligned}
\end{equation}
where ${\bf {\Lambda}}_i = diag\{1, exp(-j\frac{2\pi i}{MN}), \cdots, exp(-j\frac{2(L-1)\pi i}{MN})\}$. The channel estimates obtained by SP-DD and SP-DD-D can be regarded as approximations of the least squares estimates, and ${\bf\hat c}_{qL}$ can be computed as
\begin{equation}
{\bf{\hat c}}_{qL} = \sqrt{\frac{\beta}{\rho_F}}({\bf A}_{qp}^{H}{\bf A}_{qp})^{-1}{\bf A}_{qp}^{H}{\bf{d}}_{qF;0}^{init},
\label{eq:c_hat_qL}
\end{equation}
where ${\bf A}_{qp} = diag\{{\bf{x}}_{F;sp}\}{\bf F}_{P\times L}$.
Let ${\bf \tilde c}_{qL} = {\bf \hat c}_{qL} - {\bf c}_{qL}$. From (\ref{eq:x_F_init1_0}) and (\ref{eq:c_hat_qL}), we have
\begin{equation}
{\bf \tilde c}_{qL} = \sqrt{\frac{\beta}{\rho_F}}({\bf A}_{qp}^{H}{\bf A}_{qp})^{-1}{\bf A}_{qp}^{H}{\bf \tilde w}_{q;0},
\label{eq:c_tilde_qL}
\end{equation}
where ${\bf \tilde w}_{q;0}$ is Gaussian noise with a mean of 0 and a variance of $\delta_{w}^2 + (1-\rho_F)\kappa_q$.
\subsection{Determining Pilot Power Ratio and Power Concentration Factor}\label{OPPR}
Similar to \cite{Mishra2017} and \cite{TCLiu}, the pilot power factor can obtained through maximizing the signal-to-interference-plus-noise ratio (SINR) in the first iteration.
To this end, we need to derive the SINR. The received signal after removing pilot interference can be represented as

\begin{equation}
\begin{aligned}
{\bf{\tilde d}}_{qF;0}^{init} &=\sqrt{\frac{1 - \rho_F}{\beta}}{\bf A}_{qd0}{\bf \hat c}_{qL} \\
&+ (\sqrt{\frac{\rho_F}{\beta}}{\bf A}_{qp} + \sqrt{\frac{1 - \rho_F}{\beta}}{\bf A}_{qd0}){\bf \tilde c}_{qL} + {\bf w}_{q;0},
\label{eq:d_tqF_init1_0}
\end{aligned}
\end{equation}
\begin{equation}
\begin{aligned}
{\bf{\tilde d}}_{qF;i}^{init} = \sqrt{\frac{1}{\beta}}{\bf A}_{qdi}{\bf \hat c}_{qL} + \sqrt{\frac{1}{\beta}}{\bf A}_{qdi}{\bf \tilde c}_{qL} + {\bf w}_{q;i},
\label{eq:d_tqF_init1_i}
\end{aligned}
\end{equation}
where ${\bf A}_{qdi} = diag\{{\bf{x}}_{F;d;i}\}{\bf F}_{P\times L}{\bf {\Lambda}}_i$. Then, the data power can be computed as
\begin{equation}
\begin{aligned}
{\rm SINR}_{d;q;0} = \frac{1 - \rho_F}{\beta}E\{{\bf \hat c}_{qL}^{H}{\bf A}_{qd0}^{H}{\bf A}_{qd0}{\bf \hat c}_{qL}\},
\label{eq:SNIR_data0}
\end{aligned}
\end{equation}
\begin{equation}
\begin{aligned}
{\rm SINR}_{d;q;i} = \frac{1}{\beta}E\{{\bf \hat c}_{qL}^{H}{\bf A}_{qdi}^{H}{\bf A}_{qdi}{\bf \hat c}_{qL}\},
\label{eq:SNIR_datai}
\end{aligned}
\end{equation}
where
\begin{equation}
\begin{aligned}
E\{{\bf \hat c}_{qL}^{H}{\bf A}_{qdi}^{H}{\bf A}_{qdi}{\bf \hat c}_{qL}\} &= {\rm Tr}\{E\{{\bf c}_{qL}{\bf c}_{qL}^{H}{\bf A}_{qdi}^{H}{\bf A}_{qdi}\}\} \\
&+ {\rm Tr}\{E\{{\bf \tilde c}_{qL}{\bf \tilde c}_{qL}^{H}{\bf A}_{qdi}^{H}{\bf A}_{qdi}\}\}
\label{eq:Ehchc}
\end{aligned}
\end{equation}
According to \cite{TCLiu}, ${\rm Tr}\{E\{{\bf \tilde c}_{qL}{\bf \tilde c}_{qL}^{H}{\bf A}_{qdi}^{H}{\bf A}_{qdi}\}\}$ can be approximated as $\frac{\beta}{\rho_F}(\delta_{w}^2 + (1 - \rho)\kappa_q)L$, and ${\rm Tr}\{E\{{\bf c}_{qL}{\bf  c}_{qL}^{H}{\bf A}_{qdi}^{H}{\bf A}_{qdi}\}\}$ can be approximated as $\frac{\kappa_q}{\beta} {\rm Tr}\{{\bf R}_{hh}\}$, where ${\bf R}_{hh}$ represents channel correlation matrix. Let $s = {\rm Tr}\{{\bf R}_{hh}\}$. Then, (\ref{eq:SNIR_data0}) and (\ref{eq:SNIR_datai}) can be rewritten as
\begin{equation}
\begin{aligned}
{\rm SINR}_{d;q;0} = \frac{1 - \rho_F}{\beta^2 \rho_F}\left(\left(\delta^2_{w}+ (1 - \rho_F)\kappa_q \right)\beta^2 L + \rho_F \kappa_q s \right),
\label{eq:SNIR_data00}
\end{aligned}
\end{equation}
\begin{equation}
\begin{aligned}
{\rm SINR}_{d;q;i} = \frac{1}{\beta^2 \rho_F}\left(\left(\delta^2_{w}+ (1 - \rho_F)\kappa_q \right)\beta^2 L + \rho_F \kappa_q s \right).
\label{eq:SNIR_data11}
\end{aligned}
\end{equation}
Similarly, the ${\rm SNIR}_{n;q;i}$ can be expressed as
\begin{equation}
\begin{aligned}
{\rm SINR}_{n;q;i} =  P\delta^2_{w} + \frac{1}{\rho_F}(\delta^2_{w} + (1 - \rho)\kappa_q)L.
\label{eq:SNIR_IN}
\end{aligned}
\end{equation}

Then, the combine ${\rm SINR}$ across all subcarries can be expressed as
\begin{equation}
\begin{aligned}
{\rm SINR}_{} &\approx \frac{\rho_F^2d_1 - \rho_F d_2 + \left(Q\delta_w^2 + 1\right)\beta^3 L}{\rho_F d_3 + \left(Q\delta_w^2 + 1\right)\beta^2 L},
\label{eq:SNIR_init_beta}
\end{aligned}
\end{equation}
where $d_1 = \beta^2 L - {\rm Tr}\{{\bf R}_{hh}\}$, $d_2 = \beta^2 LQ \delta_w^2 + \beta^2 L + \beta^3 L - \beta {\rm Tr}\{{\bf R}_{hh}\}$ and $d_3 = \beta QMN\delta_w^2 - \beta^2 L$. In this paper, we set $\frac{MN}{\beta} \geq 100L$.
In addition, $Q \geq 4 \lceil N f_{max}\ {\Delta f} \rceil + 1$, where $f_{max}$ and ${\Delta f}$ represent the maximum frequency shift and carrier spacing, respectively \cite{Liu2022}. To reduce the number of ${\bf c}$ that need to be estimated, the order of BEM is set initially to $2 \lceil N f_{max}\ {\Delta f} \rceil + 1$, and after one iteration, $4 \lceil N f_{max}\ {\Delta f} \rceil + 1$.
Therefore, in (\ref{eq:SNIR_init_beta}), the $Q$ value is a constant, and only $\rho_F$ and $\beta$ are variables. Similar to \cite{Mishra2017} and \cite{TCLiu}, we take the derivative of ${\rm SINR}$ and obtain the optimal value of $\rho_F$ and $\beta$. Firstly, we set $\beta = 1$ and the derivative of (\ref{eq:SNIR_init_beta}) with respect to $\rho_F$ can be obtained, then the optimal $\rho_{Fo}$ can be expressed as

\begin{equation}
\begin{aligned}
\rho_{Fo} = \frac{-n_2 \pm \sqrt{n_2^2 + 4 n_1n_3}}{2n_1},
\label{eq:opt_rho}
\end{aligned}
\end{equation}
where $n_1 = d_1d_3$, $n_2 = 2\left(Q\delta_w^2 + 1\right)\beta^2 Ld_1$, and $n_3 = \left(Q\delta_w^2 + 1\right)\left(d_2 + \beta d_3\right)\beta^2 L$. Then, we increase the value of $\beta$ and repeat the calculation of (\ref{eq:opt_rho}). Finally, stop the calculation when one of the following conditions is satisfied: ${\rm SINR}\left(\beta, \rho_{Fo;\beta} \right) < {\rm SINR}\left(\beta-1, \rho_{Fo;\beta-1} \right)$, $\rho_{Fo;\beta}$ is not a real number, $\rho_{Fo;\beta} \not\in \left(0, 1\right)$, or reaching the maximum value of $\beta$. It is noted that (\ref{eq:opt_rho}) is derived based on two assumptions, so $\rho_F$ in (\ref{eq:opt_rho}) is approximately optimal.

\begin{table*}[htbp]
 \caption{\label{tblccp}Computational Complexity of the Proposed and Existing SP Receivers}
 \begin{center}{
 \begin{tabular}{l|c|c|c}
  \hline \textbf{Item} & \textbf{SP-DD / SP-DD-D} & \textbf{SP-BEM-MP} & \textbf{SP-aided-MP} \\
  \hline \textbf{Channel Estimation} & -- & $\mathcal{O}(Q^2L^2MNI_1)$ & $\mathcal{O}((2L^2MN + L^2 + L)I_2)$ \\
  \hline \textbf{Pilot Interference Cancellation} & -- &$\mathcal{O}((M^2N^2 + MN)I_1)$  & $\mathcal{O}((M^2N^2 + MN)I_2)$  \\
  \hline \textbf{Symbol Detection}& -- & $\mathcal{O}(N^2MI_{MP}LSI_1)$ & $\mathcal{O}(N^2MI_{MP}LSI_2)$ \\
  \hline \textbf{Total}  & $\mathcal{O}(QMNlog(MN)I_{DD})$ & \makecell{$\mathcal{O}(N^2MI_{MP}LSI_1)$, $I_{MP}LS \geq M$ \\ or $\mathcal{O}((M^2N^2 + MN)I_1)$,  otherwise} & \makecell{$\mathcal{O}(N^2MI_{MP}LSI_2)$, $I_{MP}LS \geq M$ \\ or $\mathcal{O}((M^2N^2 + MN)I)$, otherwise}\\
 \hline
 \end{tabular}
 }
 \end{center}
\end{table*}

\subsection{Complexity Analysis}\label{CMA}
It can be seen that there is no matrix inversion involved, and the computational complexity is dominated by matrix-vector products, such as (\ref{eq:bar_xTF_back}),  (\ref{eq:bar_xT_for}), etc. Fortunately, these matrix-vector products can be implemented with FFT/IFFT. Therefore, the complexity per iteration is $\mathcal{O}(QMNlog(MN))$.
Table \ref{tblccp} compares the complexity of the proposed receiver and two existing receivers, i.e., the SP-aided-MP receiver \cite{Mishra2017} and the SP-BEM-MP receiver \cite{Liu2022a}. In Table \ref{tblccp}, $I_{DD}$, $I_{MP}$, $I_{1}$ and $I_2$ represent iterations of SP-DD/SP-DD-D receiver, MP detector, SP-BEM-MP receiver, and SP-aided-MP receiver, respectively, and $S$ represents the mapping order. We can see that the computational complexity of the SP-DD/SP-DD-D receiver is lower than that of SP-BEM-MP receiver and SP-aided-MP receiver.

\section{Simulation Results}\label{SR}
In this section, we evaluate the performance of the proposed SP-DD receiver and SP-DD-D receiver under the 5G TDL-C channel \cite{3GPP}, which has a length of $L = 5$. The vehicle speeds $v$ are set to 125 km/h or 500 km/h. Moreover, due to the low effective SNR of the pilot during the the first iteration, a higher BEM order can result in poor performance in channel estimation. To solve this problem, we adopt the method in \cite{Liu2022, Liu2022a}, where a lower BEM order is used to reduce unknown parameters of the channel in first iteration, and in subsequent iterations, the BEM order is increased to ensure the accuracy of channel estimation. The order of GCE-BEM is set as the initial order $Q=5$, and after one iteration, $Q=9$. Table \ref{tbl1} summarizes other simulation parameters.
\begin{table}[htbp]
 \caption{\label{tbl1}Simulation parameters}
 \begin{center}
 \begin{tabular}{l|l}
  \hline {\bf{simulation parameter}} & {\bf{Value}}\\
  \hline Carrier frequency ($f_s$) & 4 GHz\\
  \hline Subcarrier spacing (${\Delta}f$) & 15 KHz\\
  \hline Number of delay bins ($M$) & 128\\
  \hline Number of Doppler bins ($N$) & 16\\
  \hline Modulation scheme & QPSK\\
  \hline Monte-Carlo number ($N_{MC}$) & 300\\
 \hline
 \end{tabular}
 \end{center}
\end{table}
\subsection{Convergence Analysis}
\begin{figure}[htbp]
	\centering
        \subfigure[]{\includegraphics[scale=.42]{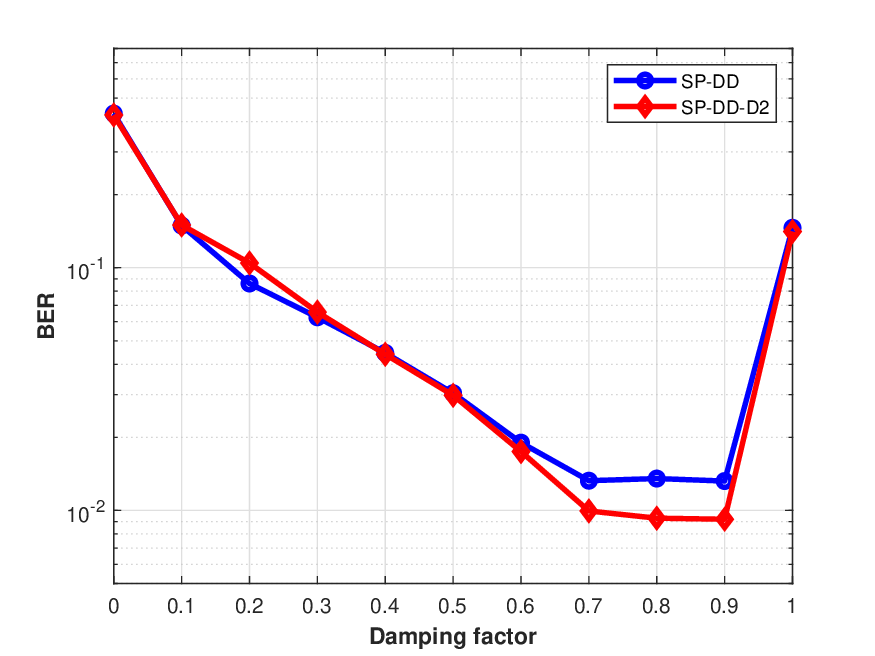}\label{FIG:DF_BER:a}}
	\centering
		\subfigure[]{\includegraphics[scale=.42]{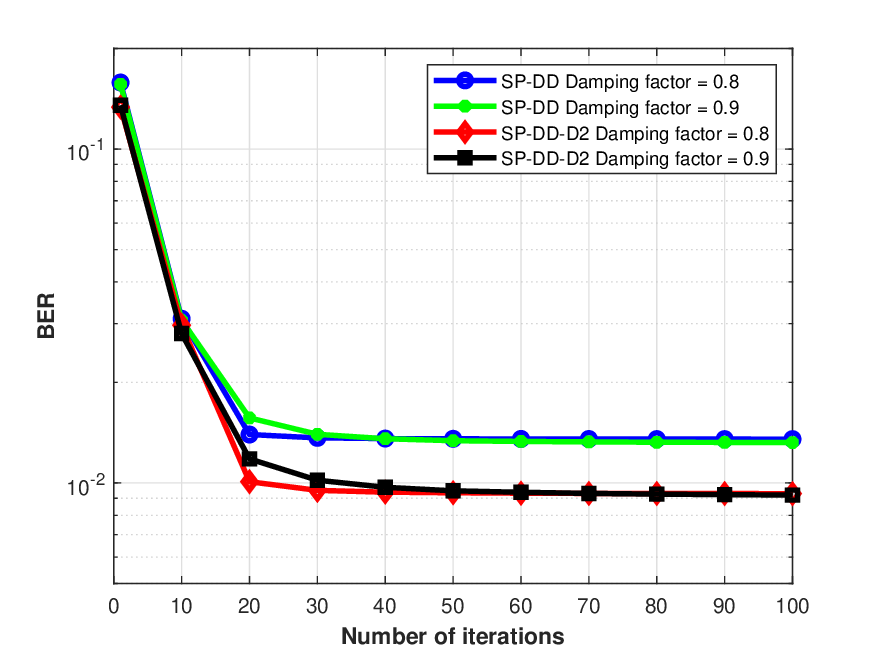}\label{FIG:DF_BER:b}}\\
	\caption{BER of the proposed SP-DD and SP-DD-D receivers versus (a) damping factor; (b) the number of iterations.
	\label{FIG:DF_BER}}
\end{figure}

In this section, we investigated two key parameters, damping factor and iteration number, for the proposed SP-DD receiver and SP-DD-D receiver. In our simulations in this subsection, the maximum number of iterations, vehicle speed $v$, $\rho_F$ and SNR are set to 100, 125 km/h,  0.2 and 12 dB, respectively. Moreover, BER is used as the performance metric. In the figures, ``SP-DD-D\#'' refers to the SP-DD-D with $\beta = \#$, and ``SP-DD" represents the SP-DD receiver. We show the BER curves under different damping factors in Fig. \ref{FIG:DF_BER}.

It can be seen from Fig. \ref{FIG:DF_BER:a} that, with a proper value of the damping factor, better BER performance can be achieved. In the range [0.7, 0.9], the system performance is not sensitive to the damping factor.Fig. \ref{FIG:DF_BER:b} shows that the receiver converges when the  number of iterations exceeds 70. So we choose 0.8 for the damping factor and 70 as the number of iterations.

\subsection{Pilot Power Allocation}
Table \ref{tblppa} summarizes the derived optimal superimposed pilot power ratio in frequency domain. It can be seen that $\rho_F$ changes with SNR. Moreover, $\rho_{Fo}$ in (\ref{eq:opt_rho}) may not be within the range of 0 to 1. Therefore, in Table \ref{tblppa}, ``--" is used to indicate this situation. In addition, $\{\cdot\}^\dagger$ represents the optimal $\rho_F$ under the optimal $\beta$.
\begin{table}[htbp]
 \caption{\label{tblppa}Optimal Pilot Power Allocation Ratio in Frequency Domain}
 \begin{center}
 \begin{tabular}{l|l|l|l|l}
  \hline SNR(dB) & 12 & 13 & 14 & 15\\
  \hline SP-DD/SP-DD-D1 & $7.41\%$ & $8.44\%$ & $9.53\%$ & $10.70\%$\\
  \hline SP-DD-D2 & ${10.41\%}^\dagger$ & ${13.44\%}^\dagger$ & ${16.27\%}^\dagger$ & ${19.08\%}$ \\
  \hline SP-DD-D4  & -- & -- & -- & ${26.85\%}^\dagger$ \\
  \hline SP-DD-D8  & -- & -- & --& -- \\
 \hline
 \end{tabular}
 \end{center}
\end{table}

\begin{figure}[htbp]
	\centering
        \subfigure[]{\includegraphics[scale=.42]{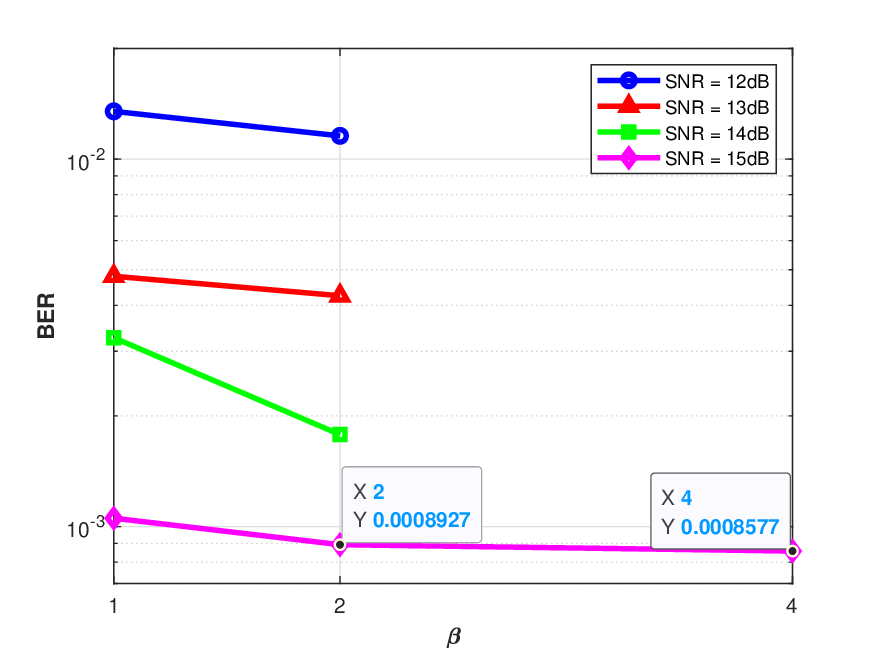}\label{FIG:Pilot_p:a}}
        \subfigure[]{\includegraphics[scale=.42]{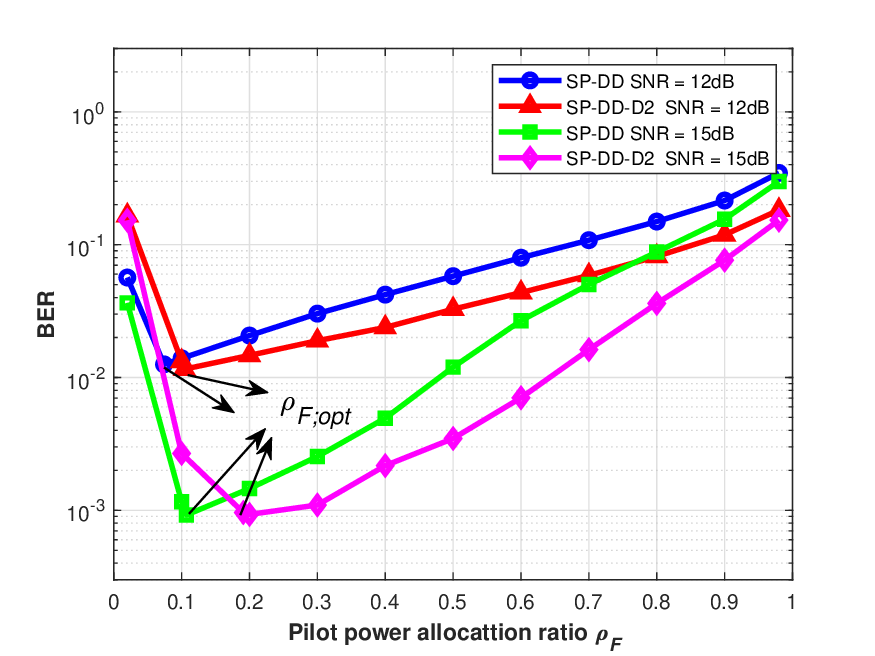}\label{FIG:Pilot_p:b}}
	\caption{BER of the proposed SP-DD and SP-DD-D receivers versus (a) $\beta$; (b) $\rho_F$.
	\label{FIG:Pilot_p}}
\end{figure}

Fig. \ref{FIG:Pilot_p} shows the BER performance of the proposed SP-DD receiver and SP-DD-D receiver versus $\beta$ and ${{\rho}}_F$. The results in Fig. \ref{FIG:Pilot_p:a} exhibit high consistency with the theoretical results in Table \ref{tblppa}. Moreover, at SNR = 15 dB, there is not much difference in BER performance between $\beta = 2$ and $\beta = 4$, while in other cases, $\beta = 2$ is optimal. Therefore, we set $\beta$ of the SP-DD-D receiver to 2.
It is noted that, when $\rho_F$ is close to 0, the initial channel estimation performance is poor due to the insufficient SP power, which in turn cannot produce good data symbol estimates and thus the CE does not improve with iterations. On the other hand, when $\rho_F$ is close to 1, most of the power will be allocated to the pilot, resulting in a lower effective SINR for the data symbols and thus poor BER performance. In addition, it can be seen that the simulation results in Fig. \ref{FIG:Pilot_p:b} show high consistency with the theoretical derivation in (\ref{eq:opt_rho}). Although $\rho_{Fo}$ varies at different SNR, in practical applications, the value of $\rho_F$ is usually pre-set as a fixed value. Comparing the results under all SNR, in the following simulation, we chose $\rho_F = 10.70\%$ for SP-DD receiver and $\rho_F = 19.08\% $ for SP-DD-D2 receiver. In addition, according to the previous settings, the power factor $\rho$ of the SP-DD-D2 receiver is about 1.2\% smaller than that of the SP-DD receiver. Moreover, for SP-DD receiver, when $\rho_F = 10.70\%$, the PAPR is 7.52 dB, while for SP-DD-D2 receiver, when $\rho_F = 19.08\%$, the PAPR is 7.23 dB. Therefore, compared to the PAPR of the SP-DD receiver, the PAPR of the SP-DD-D2 receiver is reduced by about 4\%.

\subsection{Comparison with Existing Receiver}
\begin{figure}[htbp]
	\centering
        \subfigure[]{\includegraphics[scale=.42]{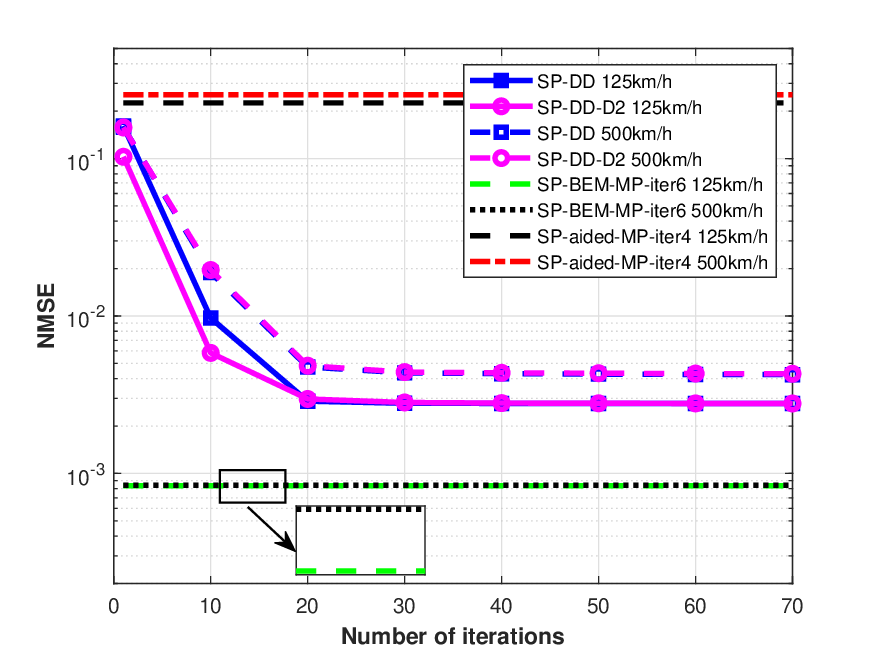}\label{FIG:MSE:a}}
	\centering
		\subfigure[]{\includegraphics[scale=.42]{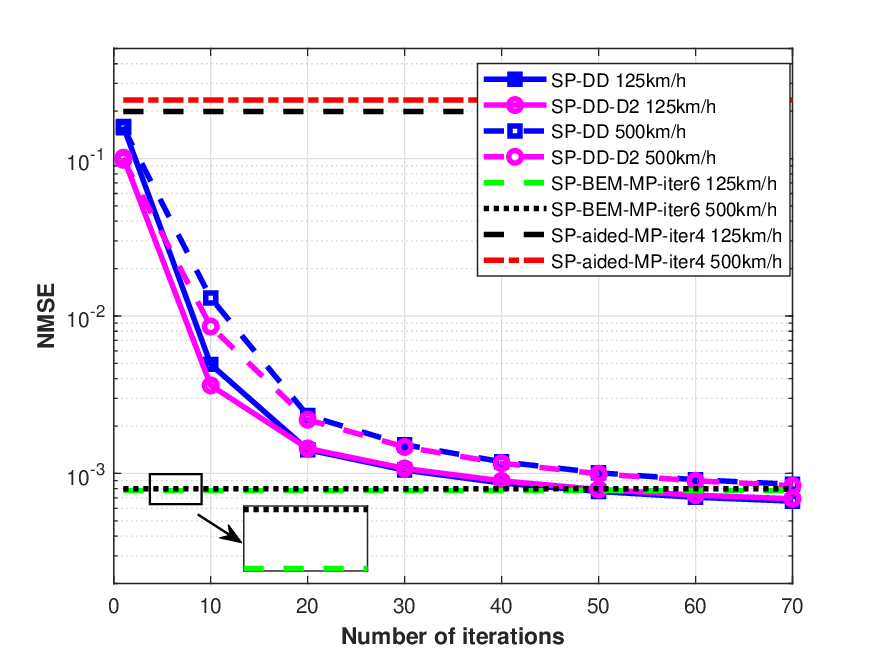}\label{FIG:MSE:b}}\\

	\caption{NMSE curves of the proposed SP-DD and SP-DD-D against the number of iteration at: (a) SNR = 12 dB, (b) SNR = 15 dB.
	\label{FIG:MSE}}
\end{figure}

In this section, we investigate the normalized mean square error (NMSE) of CE and BER performance of the proposed SP-DD receiver and SP-DD-D receiver. The MSE of CE is defined as
\begin{equation}
{\rm NMSE}(k) = \frac{1}{N_{MC}} \sum_k \frac{\|{\bf{H}} - {\bf{\hat H}}^k\|^{2}_{2}}{MNL},
\label{eq:MSE_CE}
\end{equation}
where $k$ is the iteration index in SP-DD receiver and SP-DD-D receiver, and $N_{MC}$ is the number of Monte Carlo trials. We show the NMSE curves under different vehicle speed in Fig. \ref{FIG:MSE}. Moreover, the NMSE value of the SP-BEM-MP algorithm and the NMSE value of the SP-aided-MP algorithm are selected as benchmarks.

From Fig. \ref{FIG:MSE}, the CE accuracy of the SP-aided-MP receiver is the worst due to the presence of Doppler spread in the channel, and the NMSE performance of SP-BEM-MP receiver is the best among all receivers. In addition, when the SNR is high, the main factor affecting NMSE is the BEM model error. Therefore, in Fig. \ref{FIG:MSE}, the NMSE of the SP-BEM-MP receiver is almost the same under different SNRs. Moreover, it can be seen that at the beginning of the iterative process, the NMSE of the SP-DD-D2 receiver is lower than that of the SP-DD receiver, which is consistent with the analysis in Section IV-B. After the algorithm converges, the channel estimation accuracy of the two receivers is almost the same. In addition, when SNR = 12dB, the NMSE of the SP-DD/SP-DD-D receiver is significantly different from that of SP-BEM-MP receiver. Moreover, when the SNR is high, the NMSE of the SP-DD/SP-DD-D receiver is significantly improved, as shown in Fig. \ref{FIG:MSE:b}.

\begin{figure}[htbp]
	\centering
        \subfigure[]{\includegraphics[scale=.42]{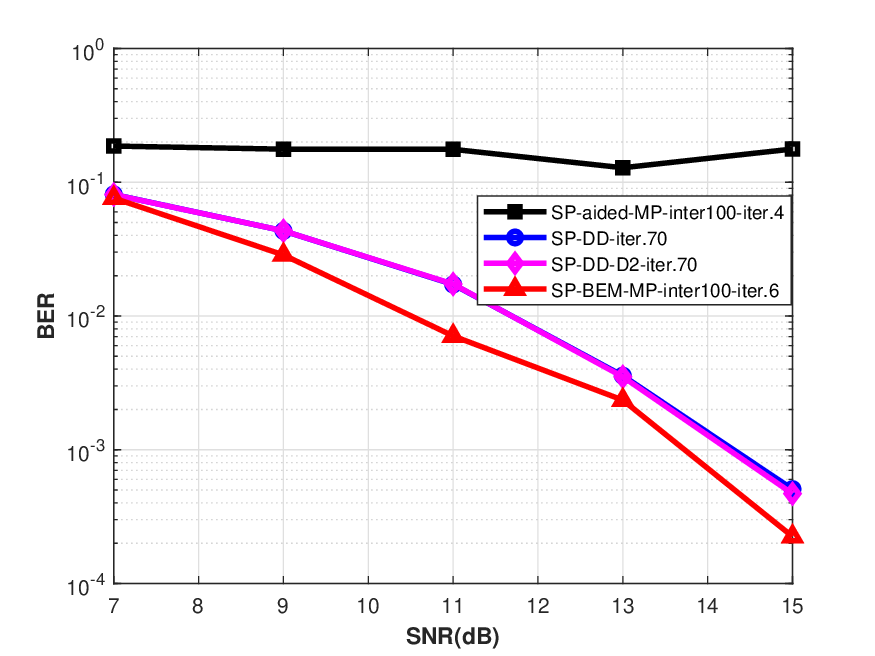}\label{FIG:Compare:a}}
	\centering
		\subfigure[]{\includegraphics[scale=.42]{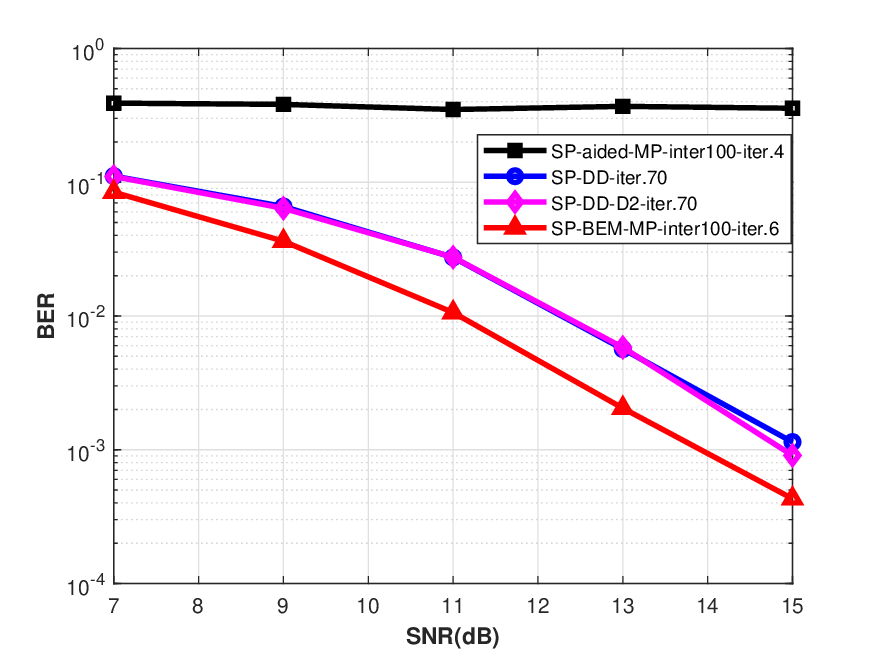}\label{FIG:Compare:b}}
	\caption{BER performance of OTFS receivers in different scenarios: (a) $v = 125$ km/h, (b) $v = 500$ km/h.
	\label{FIG:Compare}}
\end{figure}

Next, we evaluate the BER performance of the proposed scheme, and Fig. \ref{FIG:Compare} shows the BER performance of the four OTFS receivers, i.e.,  SP-DD receiver, SP-DD-D receiver, SP-BEM-SP receiver and SP-aided-SP receiver. For these four schemes, we compare their performance after the convergence. After convergence, the BER performance of the SP-BEM-MP receiver is the best, followed by the SP-DD-D2 receiver and SP-DD receiver, and the performance of the SP-aided-MP receiver is the worst. When BER=$10^{-3}$ and the speed is 500 km/h, the performance difference between the SP-BEM-MP receiver and the SP-DD/SP-DD-D2 is about 1 dB, while the difference between the SP-BEM-MP receiver and the SP-DD/SP-DD-D2 receiver is less than 0.6 dB, when BER=$10^{-3}$ and the speed is 125 km/h. From Fig. \ref{FIG:Compare}, we can see that the BER of the SP-DD-D2 receiver is slightly lower than that of the SP-DD receiver. To sum up, compared with the SP-DD receiver, the SP-DD-D receiver has slightly better BER performance, with the pilot power reduced by 1.2\% and the PAPR of the signal dropped by 4\%.

\subsection{Runtime Comparison}\label{RTC}
We then evaluate the computational complexity of the SP-DD-D scheme using the average runtime. The parameter settings in the simulation are as follows: $M = 128$, $N = 16$, $Q = 9$, $L = 5$, $I_{DD} = 60 $, $I_{MP} = 100$, $I_1 = 6$ and $I_2 = 4$. Table \ref{tblccpa} shows the average runtime of various algorithms when SNR = 15dB. We can see that the computational complexity of the proposed SP-DD/SP-DD-D receiver is significantly lower than that of existing SP-OTFS receivers.

\begin{table}[htbp]
 \caption{\label{tblccpa}Runtime of the Proposed and the Existing SP Receiver}
 \begin{center}
 \begin{tabular}{c|c|c|c|c}
  \hline \textbf{speed(km/h)} & \textbf{SP-DD-D2}  &\textbf{SP-DD} & \textbf{SP-BEM-MP} & \textbf{SP-aided-MP} \\
  \hline {125 } & 0.33 s  &0.34 s & 28.95 s & 15.81 s \\
  \hline {500 } & 0.33 s  &0.36 s & 39.08 s & 15.63 s\\
 \hline
 \end{tabular}
 \end{center}
\end{table}

\begin{figure}[htbp]
	\centering
        \subfigure[]{\includegraphics[scale=.42]{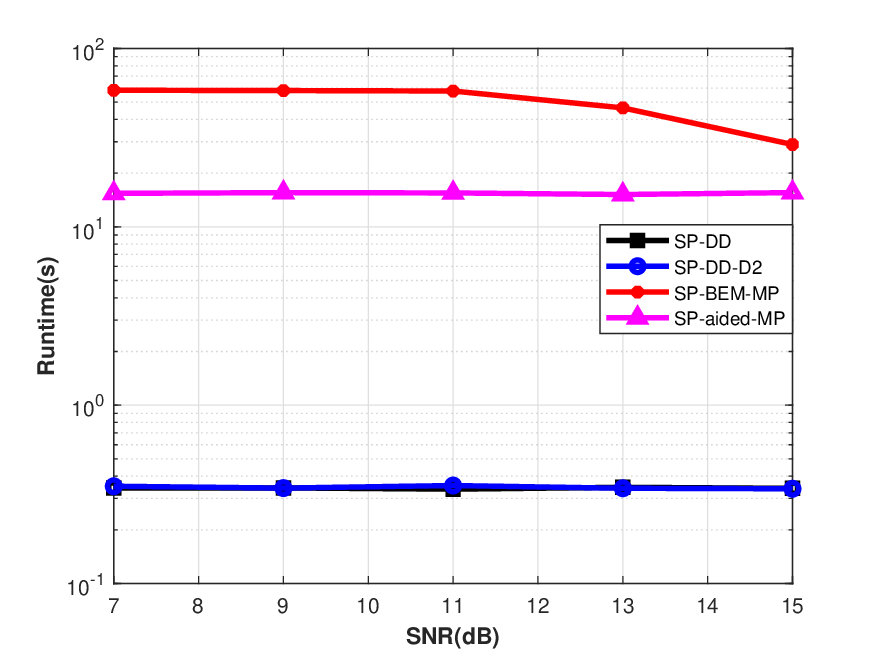}\label{FIG:Runtime:a}}
	\centering
		\subfigure[]{\includegraphics[scale=.42]{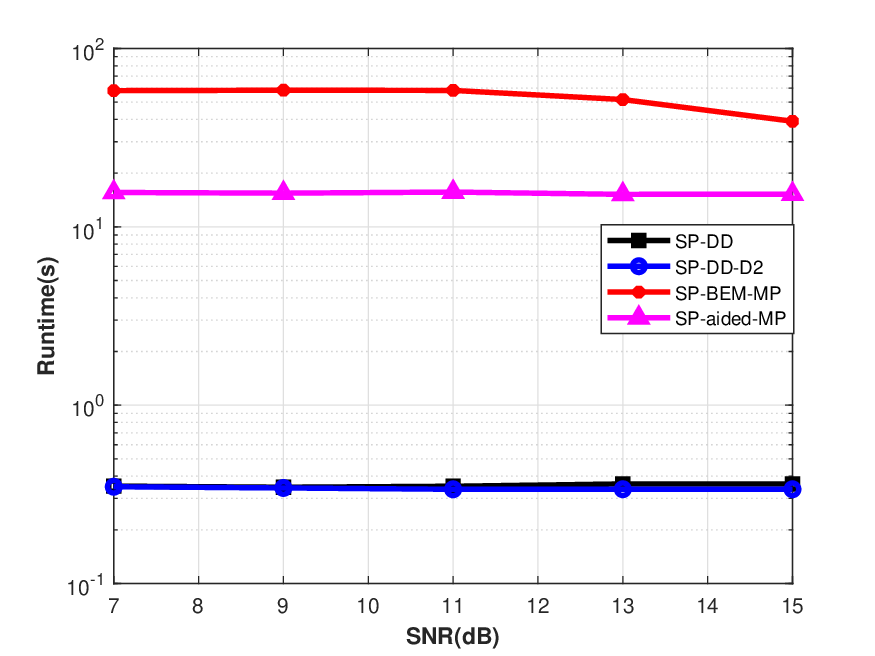}\label{FIG:Runtime:b}}
	\caption{Average runtime of OTFS receivers in different scenarios: (a) $v = 125$ km/h, (b) $v = 500$ km/h.
	\label{FIG:Runtime}}
\end{figure}

Fig. \ref{FIG:Runtime} compares the average runtime of the SP-DD receiver, the SP-DD-D2 receiver, SP-BEM-MP receiver, and the SP-aided-MP receiver. The results are obtained using MATLAB (R2020a) on a computer with a 6-core Intel i7-8700 processor.
According to Fig. \ref{FIG:Runtime}, we can see that that the proposed SP-DD receiver and SP-DD-D2 receiver are much faster than the SP-BEM-MP receiver and SP-aided-MP receiver. Moreover, according to Table \ref{tblccpa}, when SNR = 15dB, the runtime of the SP-BEM-MP receiver is 103.07 times longer than that of the SP-DD-D2 receiver. To sum up, the SP-DD-D2 receiver has the lowest complexity among all receivers.
According to the results in Fig. \ref{FIG:Compare}, Fig. \ref{FIG:Runtime} and Table \ref{tblccpa}, the proposed SP-DD/SP-DD-D2 scheme greatly reduces the computational complexity while with only a slight loss in the BER performance, compared to the SP-BEM-MP receiver.

\section{Conclusion} \label{sec7}

In this paper, we have investigated the issue of joint channel estimation and signal detection in OTFS systems with SP to achieve high transmission efficiency. To address the high computational complexity issue of the existing receiver, we proposed a new receiver called the SP-DD receiver, leveraging the message-passing techniques. It is shown that the SP-DD receiver delivers similar performance with drastically reduced complexity. To facilitate CE in the proposed receiver, the pilot signal is designed to achieve pilot power concentration in the frequency domain, leading to the SP-DD-D receiver. It is shown that SP-DD-D can reduce the power of the pilot signal and the PAPR of the signal. Extensive simulation results demonstrate the superiority of the proposed receiver.

\bibliographystyle{IEEEtran}
\bibliography{myreferences}




%

\end{document}